\documentclass[twocolumn,twocolappendix]{aastex631}
\usepackage{amsmath}
\usepackage{CJKutf8}

\begin{document}
\title{Stellar Obliquity of the Ultra-Short-Period Planet System HD~93963}

\author[0000-0003-3860-6297]{Huan-Yu Teng 
\begin{CJK*}{UTF8}{gbsn}
(滕环宇)
\end{CJK*}} 
\affiliation{Korea Astronomy and Space Science Institute
776 Daedeok-daero, Yuseong-gu, Daejeon 34055, Republic of Korea}
\affiliation{Institute for Astronomy, University of Hawai`i, 2680 Woodlawn Drive, Honolulu, HI 96822 USA}
\affiliation{CAS Key Laboratory of Optical Astronomy, National Astronomical Observatories, Chinese Academy of Sciences, Beijing 100101, China}

\author[0000-0002-8958-0683]{Fei Dai
\begin{CJK*}{UTF8}{gbsn}
(戴飞)
\end{CJK*}} 
\affiliation{Institute for Astronomy, University of Hawai`i, 2680 Woodlawn Drive, Honolulu, HI 96822 USA}

\author[0000-0001-8638-0320]{Andrew W. Howard} 
\affiliation{Department of Astronomy, California Institute of Technology, Pasadena, CA 91125, USA}

\author[0000-0003-1312-9391]{Samuel Halverson} 
\affiliation{Jet Propulsion Laboratory, California Institute of Technology, 4800 Oak Grove Drive, Pasadena, CA 91109, USA} 

\author[0000-0002-0531-1073]{Howard Isaacson} 
\affiliation{{Department of Astronomy,  University of California Berkeley, Berkeley CA 94720, USA}}
\affiliation{Centre for Astrophysics, University of Southern Queensland, Toowoomba, QLD, Australia}

\author[0000-0002-5486-7828]{Eiichiro Kokubo
\begin{CJK*}{UTF8}{gbsn}
(小久保英一郎)
\end{CJK*}} 
\affiliation{National Astronomical Observatory of Japan, 2-21-1 Osawa, Mitaka, Tokyo 181-8588, Japan } 

\author[0000-0003-3856-3143]{Ryan A. Rubenzahl}  
\affiliation{Center for Computational Astrophysics, Flatiron Institute, 162 Fifth Avenue, New York, NY 10010, USA} 

\author[0000-0003-3504-5316]{Benjamin Fulton} 
\affiliation{NASA Exoplanet Science Institute/Caltech-IPAC, MC 314-6, 1200 E California Blvd, Pasadena, CA 91125, USA} 

\author[0000-0002-5812-3236]{Aaron Householder} 
\affiliation{Department of Earth, Atmospheric and Planetary Sciences, Massachusetts Institute of Technology, Cambridge, MA 02139, USA}
\affil{Kavli Institute for Astrophysics and Space Research, Massachusetts Institute of Technology, Cambridge, MA 02139, USA}

\author[0000-0001-8342-7736]{Jack Lubin} 
\affiliation{Department of Physics \& Astronomy, University of California Los Angeles, Los Angeles, CA 90095, USA} 

\author[0000-0002-8965-3969]{Steven Giacalone} 
\affiliation{Department of Astronomy, California Institute of Technology, Pasadena, CA 91125, USA} 

\author[0000-0002-9305-5101]{Luke Handley} 
\affiliation{Department of Astronomy, California Institute of Technology, Pasadena, CA 91125, USA} 

\author[0000-0002-4290-6826]{Judah Van Zandt} 
\affil{Department of Physics \& Astronomy, University of California Los Angeles, Los Angeles, CA 90095, USA} 

\author[0000-0003-0967-2893]{Erik A. Petigura} 
\affiliation{Department of Physics \& Astronomy, University of California Los Angeles, Los Angeles, CA 90095, USA}

\author[0000-0001-7664-648X]{J. M. Joel Ong \begin{CJK*}{UTF8}{gbsn}
(王加冕)
\end{CJK*}} 
\altaffiliation{NASA Hubble Fellow} 
\affiliation{Institute for Astronomy, University of Hawai`i, 2680 Woodlawn Drive, Honolulu, HI 96822 USA}

\author[0000-0001-5728-4735]{Pranav Premnath} 
\affiliation{Department of Physics \& Astronomy, University of California Irvine, Irvine, CA 92697, USA} 

\author[0000-0002-0971-6078]{Haochuan Yu \begin{CJK*}{UTF8}{gbsn}
(于皓川)
\end{CJK*}} 
\affiliation{Sub-department of Astrophysics, Department of Physics, University of
Oxford, Oxford OX1 3RH, UK} 


\author[0009-0004-4454-6053]{Steven R. Gibson} 
\affil{Caltech Optical Observatories, Pasadena, CA, 91125, USA}

\author{Kodi Rider} 
\affil{Space Sciences Laboratory, University of California Berkeley, Berkeley, CA 94720, USA}

\author[0000-0001-8127-5775]{Arpita Roy} 
\affiliation{Astrophysics \& Space Institute, Schmidt Sciences, New York, NY 10011, USA} 

\author[0000-0002-6525-7013]{Ashley Baker} 
\affil{Caltech Optical Observatories, Pasadena, CA, 91125, USA}

\author{Jerry Edelstein} 
\affil{Space Sciences Laboratory, University of California Berkeley, Berkeley, CA 94720, USA}

\author{Chris Smith} 
\affil{Space Sciences Laboratory, University of California Berkeley, Berkeley, CA 94720, USA}

\author[0000-0002-6092-8295]{Josh Walawender} 
\affiliation{W. M. Keck Observatory, 65-1120 Mamalahoa Hwy, Waimea, HI 96743}

\author[0000-0002-2783-0755]{Byeong-Cheol Lee} 
\affiliation{Korea Astronomy and Space Science Institute 776 Daedeok-daero, Yuseong-gu, Daejeon 34055, Republic of Korea}

\author[0009-0008-3430-1027]{Yu-Juan Liu} 
\affiliation{CAS Key Laboratory of Optical Astronomy, National Astronomical Observatories, Chinese Academy of Sciences, Beijing 100101, China}

\author[0000-0002-4265-047X]{Joshua N. Winn} 
\affiliation{Department of Astrophysical Sciences, Princeton University, 4 Ivy Lane, Princeton, NJ 08544, USA} 

\begin{abstract}
We report an observation of the Rossiter-McLaughlin (RM) effect  of the transiting planet HD~93963~Ac, a mini-Neptune planet orbiting a G0-type star with an orbital period of $P_{\rm{c}} = 3.65\,\mathrm{d}$, accompanied by an inner super-Earth planet with $P_{\rm{b}} = 1.04\,\mathrm{d}$. We observed a full transit of planet~c on 2024~May~3rd~UT with Keck/KPF. The observed RM effect has an amplitude of $\sim 1\,\mathrm{m\,s}^{-1}$ and implies a sky-projected obliquity of $\lambda = 14^{+17}_{-19}$~degrees for HD~93963~Ac. Our dynamical analysis suggests that the two inner planets are likely well aligned with the stellar spin, to within a few degrees, thus allowing both to transit. Along with WASP-47, 55~Cnc, and HD~3167, HD~93963 is the fourth planetary system with an ultra-short-period planet and obliquity measurement(s) of any planet(s) in the system. HD~93963, WASP-47, and 55~Cnc favor largely coplanar orbital architectures, whereas HD~3167 has been reported to have a large mutual inclination ($\sim$100$^\circ$) between its transiting planets~b and~c. In this configuration, 
the probability that both planets transit is low. Moreover, one planet would quickly evolve to be non-transiting due to nodal precession. Future missions such as ESO/PLATO should detect the resulting transit duration variations. We encourage additional obliquity measurements of the HD~3167 system to better constrain its orbital architecture.
\end{abstract}


\clearpage\newpage
\section{Introduction} \label{sec:intro}
The true stellar obliquity $\psi$, or namely the true spin-orbit angle, is the angle between the rotation axis of the host star and the normal of a planet's orbital plane, and it is a key indicator of planet formation and migration history. The sky-projected stellar obliquity $\lambda$ can be measured through the Rossiter-McLaughlin (RM) effect \citep{Rossiter1924, McLaughlin1924}, a distortion of stellar line profiles that occurs during planetary transits. As a planet sequentially blocks the red-shifted and blue-shifted limbs of the rotating stellar surface, one measures an anomalous radial velocity (RV) variation or line shape distortion during the course of the transit. The pattern of the observed Rossiter-McLaughlin effect encodes $\lambda$. The true stellar obliquity $\psi$ -- the angle between the planet's orbital vector $\hat{n}_{\rm{orb}}$ and the star's spin vector $\hat{n}_{\star}$ -- can be calculated by combining information about $\lambda$, the stellar inclination  $i_{\star}$, and the planetary orbital inclination $i_{\rm{orb}}$ \citep[see, e.g.,][]{Albrecht2021,Masuda2020}. 

To date, over 200 planetary systems have stellar obliquity measurements\footnote{Data from TEPCat \url{https://www.astro.keele.ac.uk/jkt/tepcat/obliquity.html}}, most of which are for hot Jupiters. Only three of the stars with obliquity measurements on any of the planets in the system host ultra-short-period (USP) planets, a category of planets loosely defined as having sizes compatible with being of terrestrial composition and orbital periods shorter than about 1 day \citep{SO2014,Winn2018,Goyal}. USP planets orbit approximately 1\% of Sun-like stars. They are unlikely to have formed in their present orbits because they would have been within the inner disk edge of the protoplanetary disk, and inside of  the dust sublimation radius \citep{Isella2006,Lee2017}.

Instead, USP planets are thought to begin on longer-period orbits (a few days, comparable to the innermost planets of more typical Kepler-like systems) and subsequently migrate inward \citep{Dai2018,Petrovich2019,Pu2019,Millholland2020}. Inward migration might occur as a consequence of eccentricity or obliquity tides that are dissipated within the planet. To maintain the planet's non-zero eccentricity or obliquity, the planet must have additional companions that provide the necessary dynamical influence. In this light, it is interesting to note that USPs are almost always accompanied by nearby planets \citep{Winn2018,Dai2021,Adams2021}.

Two observational findings further support USP inward migration: (1) USPs have relatively large period ratios relative to their nearest neighboring planets \citep{Winn2018}, defying the ``peas-in-a-pod'' pattern of regular orbital spacing reported in other compact multi-planet systems \citep[e.g.,][]{Weiss2018}.  (2) While most Kepler-like multi-planet systems generally have low sky-projected mutual inclinations of $1$--$2^\circ$ \citep{Fabrycky2014}, many USPs exhibit moderate sky-projected mutual inclinations of about $5$--$15^\circ$ relative to their nearest neighbors \citep{Dai2018}. For example, K2-266 b is inclined by about $15^\circ$ compared to its outer planets \citep{Rodriguez2018}. The secular interaction scenario \citep{Petrovich2019,Pu2019} can explain these findings. In this scenario, the secular interactions within a multi-planet system launch the innermost planet into an eccentric and inclined orbit. The elevated eccentricity allows the innermost planet to migrate inward via eccentricity tides, with dissipation occurring inside the planet, similar to the case of Io and Jupiter \citep{Peale1979}.  Over time, the eccentricity is damped, but the inclined orbit persists due to slower tidal realignment, because it requires the dissipation to occur inside the host star instead of the planet.
After these events, the USPs are observed on circularized and mildly misaligned orbits (5--15$^{\circ}$, \citealt{Dai2018}). Depending on the strength of tidal dissipation or star-planet magnetic interaction \citep{Colle,Lee2025}, this inward migration journey could take many Gyr to complete. Indeed, recent results suggest that USP host stars are older than the field stars, and more likely to be thick disk members \citep{Tu,Schmidt}. Alternative scenarios that produce USP planets before disk dispersal \citep[e.g.][]{Becker2021} are inconsistent with the more mature ages of the USP host stars.

In the secular interaction framework, the outer companions to USPs should remain largely unexcited and thus they should remain on nearly aligned orbits \citep{Petrovich2019,Pu2019}. In this work, we report on an observation that checks on this prediction for a particular system. HD~93963~A (TOI-1797) is a G0 V star in a visual binary system that hosts at least two transiting planets discovered by Transiting Exoplanet Survey Satellite (TESS; \citealt{Ricker2014}) in 2022. The two small planets reside on either side of the radius valley \citep[Fulton gap, ][]{Fulton2017}. The inner and smaller planet ``b'' is a USP super-Earth with $R_{\rm{b}}=1.35\,R_{\oplus}$, $M_{\rm{b}}=4.31\,M_{\oplus}$, and $P_{\rm{b}}=1.04\,\rm{d}$, while the outer and larger planet ``c'' is a mini Neptune with $R_{\rm{c}}=3.23\,R_{\oplus}$, $M_{\rm{c}}=18.4\,M_{\oplus}$, and $P_{\rm{c}}=3.65\,\rm{d}$. Planet c has a mean density of $\rho = 3.1\,\rm{g}\,\rm{cm}^{-3}$, suggestive of a substantial H/He envelope. \citet{VanZandt2025} also found tentative evidence for a distant companion from long-term RV monitoring.

In Section \ref{sec:obs}, we describe our spectroscopic observations. In Section \ref{sec:star}, we present the derivation of the stellar properties. In Section \ref{sec:rm}, we explain our joint analysis incorporating the light curve, the RM effect, and systematic effects, and report our fitting results. In Section \ref{sec:dyn}, we provide a dynamical analysis of the HD~93963 system. In Section \ref{sec:discuss}, we discuss the obliquities of USP hosting systems with special attention to HD~3167,
the only system with a USP planet for which a very high mutual inclination between two planets has been reported. Finally, we conclude the paper in Section \ref{sec:conclusion}.

\section{Spectroscopic Observations} \label{sec:obs}

We observed the RM effect of HD~93963~Ac on 2024~May~3rd~(UTC) with the Keck Planet Finder (KPF; \citealt{Gibson2024}) on the 10m Keck~I Telescope at Mauna~Kea. The spectrograph offers a resolution of $\sim$98,000 and covers a wavelength range of 4450--8700\,\AA. Its wavelength calibration system includes a Th-Ar lamp, a laser frequency comb, a Fabry-P\'{e}rot etalon, and a solar calibrator \citep{Rubenzahl2023}. KPF achieves a noise floor of $\sim 0.3\,\mathrm{m\,s}^{-1}$.

\begin{figure*}[ht]
\centering
\includegraphics[width=0.95\linewidth]{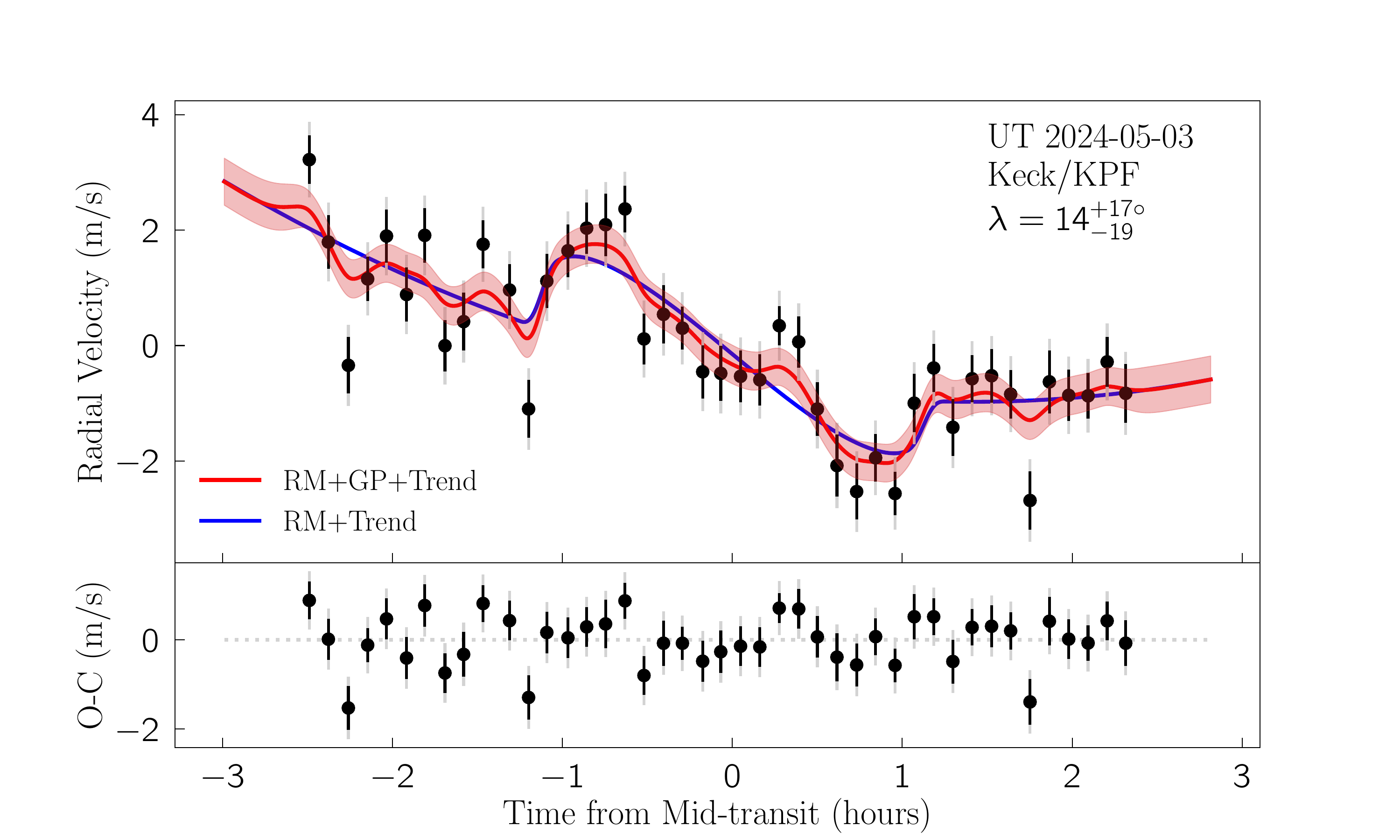}
\caption{The radial velocities (RVs) during the transit event of HD~93963~Ac on 2024 May 3rd (UTC), measured by Keck/KPF spectra. For each data point, the black errorbar is the observational uncertainty while the additional gray one indicates the extra jitter from the fitting. In the upper subplot, the solid red line is the final best-fit model ($\lambda = 14^{+17\circ}_{-19}$), incorporating Rossiter-McLaughlin (RM) effect, Gaussian Process (GP), and RV long-term trend, and the red shade shows the 1-$\sigma$ uncertainty. The solid blue line is also the final best-fit model but without GP plotted. In the lower panel, the residuals are given by subtracting the best-fit model, including RM, GP and long-term trend. \label{fig:rm}}
\end{figure*}

The binary companion is located $5.9''$ away \citep{Serrano2022}, far enough for its light to be excluded from our spectroscopic observations. We scheduled the observations according to the ephemeris provided by \citet{Serrano2022} and captured the full transit event. Observations began $\sim$1.2 hours before ingress and ended $\sim $1.1 hours after egress. We used an exposure time of 360 seconds, yielding a signal-to-noise ratio (S/N) of $\sim$270 near 5500\,\AA. In total, we obtained 43 usable spectra, 25 of which were taken during the transit. The spectra were reduced using the KPF Data Reduction Pipeline (KPF-DRP), publicly available on \href{https://github.com/Keck-DataReductionPipelines/KPF-Pipeline}{GitHub}. Our radial velocity (RV) measurements are shown in Figure \ref{fig:rm} and listed in Table \ref{tab:data}.

\begin{table}[]
\centering
\begin{tabular}{ccc}
\hline\hline 
$\rm{BJD}-2457000$ & RV $(\rm{m}\, \rm{s}^{-1})$& $\sigma_{\rm{RV}}\, (\rm{m}\,\rm{s}^{-1})$  \\
\hline 
3433.728415&3.222 &0.420 \\
3433.73309&1.794 &0.463 \\
3433.737992&-0.342 &0.488 \\
3433.742721&1.155 &0.384 \\
3433.747344&1.897 &0.458 \\
3433.752257&0.885 &0.470 \\
3433.756709&1.910 &0.474 \\
3433.761678&-0.003 &0.445 \\
3433.766252&0.415 &0.503 \\
3433.771031&1.756 &0.415 \\
3433.777514&0.963 &0.445 \\
3433.78218&-1.098 &0.500 \\
3433.786649&1.117 &0.469 \\
3433.791826&1.643 &0.456 \\
3433.796419&2.035 &0.444 \\
3433.801052&2.094 &0.542 \\
3433.80581&2.368 &0.406 \\
3433.810469&0.115 &0.439 \\
3433.815252&0.541 &0.505 \\
3433.819898&0.301 &0.371 \\
3433.824863&-0.455 &0.460 \\
3433.829297&-0.482 &0.474 \\
3433.834143&-0.532 &0.448 \\
3433.838872&-0.594 &0.447 \\
3433.843633&0.344 &0.338 \\
3433.848411&0.064 &0.444 \\
3433.852972&-1.099 &0.468 \\
3433.857782&-2.079 &0.539 \\
3433.862608&-2.530 &0.484 \\
3433.86726&-1.944 &0.412 \\
3433.872026&-2.564 &0.374 \\
3433.876723&-0.998 &0.506 \\
3433.881494&-0.388 &0.417 \\
3433.886199&-1.416 &0.492 \\
3433.890912&-0.575 &0.410 \\
3433.895688&-0.523 &0.466 \\
3433.900369&-0.842 &0.422 \\
3433.90512&-2.684 &0.509 \\
3433.909863&-0.628 &0.546 \\
3433.914601&-0.862 &0.444 \\
3433.919352&-0.871 &0.397 \\
3433.924022&-0.282 &0.433 \\
3433.928574&-0.828 &0.512 \\
\hline\hline
\end{tabular}
\caption{Keck/KPF data during transit. The airmass ranged from 1.01 and 1.26 during the observation. More information about the observing weather condition can be found at \url{http://mkwc.ifa.hawaii.edu/}. }
\label{tab:data}
\end{table}

\section{Stellar Parameters} \label{sec:star}
\begin{deluxetable}{lccc}
\tablecaption{Stellar Parameters of HD~93963~A} \label{tab:stellar_para}
\tablehead{
\colhead{Parameters (Unit)} &  \colhead{Value and Uncertainty} & \colhead{Reference}}
\startdata
TIC ID & 368435330 & A\\
R.A. & 10:51:06.51 & A\\
Dec. & +25:38:28.19 & A\\
$V$ (mag)  & 9.18 $\pm$ 0.02& A\\
$J$ (mag)  & 8.108 $\pm$ 0.034& A\\
$H$ (mag)  & 7.807 $\pm$ 0.034& A\\
$K$ (mag)  & 7.776 $\pm$ 0.024& A\\
Distance (pc)& $82.34\pm0.39$ & A\\
$T_{\text{eff}} ~(\rm{K})$ & $5947\pm100$ & B \\
$\log~g~(\text{dex})$ &$4.46 \pm 0.10$& B \\
$[\text{Fe/H}]~(\text{dex})$ &$0.10 \pm 0.06$& B \\
$v\sin\,i$ ($\rm{km}\,\rm{s}^{-1}$) &$2.93 \pm 1.00$& B$^{\dagger}$ \\
$v\sin\,i$ ($\rm{km}\,\rm{s}^{-1}$) &$5.9 \pm 0.8$& C$^{\dagger}$ \\
$M_{\star} ~(M_{\odot})$ &$1.07 \pm 0.02$& B \\
$R_{\star} ~(R_{\odot})$ &$1.03 \pm 0.02$& B \\
$P_{\rm{rot}}$ (days) & $12.4 \pm 1.4$ (TESS Photometry) & B \\
$P_{\rm{rot}}$ (days) & $12.8 \pm 1.8$ (TESS Photometry) & C \\
$P_{\rm{rot}}$ (days) & $11.02_{-0.08}^{+0.05}$ (RV+GP) & C \\
Age (Gyr) & $1.6^{+1.0}_{-0.8}$ (Isochrone)& B \\
Age (Gyr) & $1.9^{+0.5}_{-0.4}$ (Gyrochronology) & B \\
\enddata
\tablecomments{A: TICv8; B: This work; C: \citet{Serrano2022};  $^{\dagger}$These two $v\sin\,i$ results are derived from spectroscopy.}
\end{deluxetable}
\begin{figure}
\centering
\includegraphics[width=1.\linewidth]{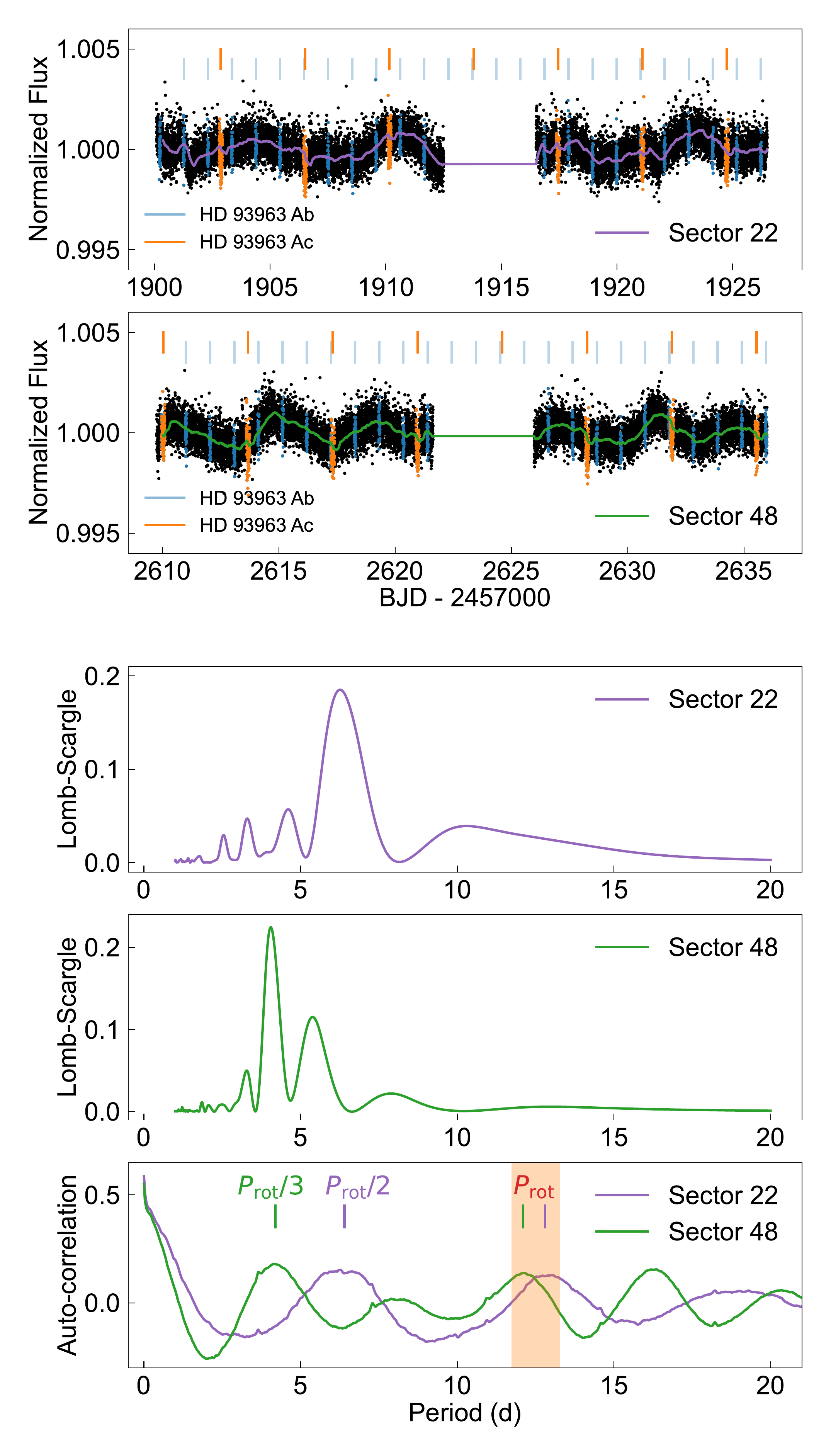}
\caption{Top two panels: {\it TESS} 2-min-cadence light curves of HD~93963~A, with the upper plot showing Sector 22 and lower plot showing Sector 48. Smoothed light curves are also plotted, to give a clearer view of the stellar variation. Transits of planets b and c are marked in blue and orange, respectively. Bottom three panels: Period analysis of the {\it TESS} light curves. The top and middle panels show the Lomb-Scargle periodograms of the light curves from Sectors 22 and 48, respectively. The bottom panel shows the auto-correlation function of the light curves from Sectors 22 and 48, with rotational periods and the harmonics marked.}
\label{fig:tess_lc}
\end{figure}

\subsection{Atmospheric Parameters}\label{sec:atm_param}
We observed HD~93963~A with Keck/HIRES in iodine-free mode on 2020~December~25th to refine the stellar parameters. The spectrum was obtained with a signal-to-noise ratio of 105 at the spectra center, under a spectral resolution of $\sim$72,000 and wavelength range of 3360--8100 \AA. The raw frames and the spectral information can be downloaded from the Keck Observatory Archive website\footnote{\url{https://koa.ipac.caltech.edu/cgi-bin/KOA/nph-KOAlogin}}. We used the \texttt{SpecMatch-Syn} pipeline\footnote{\url{https://github.com/petigura/specmatch-syn}} \citep{Petigura2017} to determine the spectroscopic parameters, including the effective temperature ($T_{\rm{eff}}$), surface gravity ($\log g$), metallicity ($\rm{[Fe/H]}$), and projected rotational velocity ($v\sin i_{\star}$). Briefly, \texttt{SpecMatch-Syn} models the observed spectra by matching them with synthetic spectra, using interpolation over a precomputed grid \citep{Coelho2005} in the parameter space of ($T_{\rm{eff}}$, $\log g$, $\rm{[Fe/H]}$, $v\sin i_{\star}$). From this procedure, we obtained $T_{\rm{eff}} = 5950 \pm 100\,\mathrm{K}$, $\log g = 4.46 \pm 0.10$, and $\rm{[Fe/H]}=0.10\pm 0.06$, consistent with the previous measurements of $T_{\rm{eff}} = 5987 \pm 64\,\mathrm{K}$, $\log g = 4.49 \pm 0.11$, and $\rm{[Fe/H]} = 0.10 \pm 0.04$ from \citet{Serrano2022}. However, our analysis yielded a projected rotational velocity of $v\sin i_{\star} = 2.93 \pm 1.00\,\mathrm{km\,s}^{-1}$, which disagrees with the previous measurement of $5.9 \pm 1.00\,\mathrm{km\,s}^{-1}$. Our new measurement of $v\sin i_{\star}$ agrees with the result from our Rossiter-McLaughlin analysis, $v\sin i_{\star} = 2.96 \pm 0.64\,\mathrm{km\,s}^{-1}$ (Section \ref{sec:rm}). The systematic uncertainty of $v\sin i$ using \texttt{SpecMatch-Syn} with Keck/HIRES spectra of FGK stars can be down to $1\, \mathrm{km\,s}^{-1}$, and the reliability of $v\sin i_{\star}$  was verified using 43 stars in \citet{Albrecht2012} by cross-checking between the spectroscopic results and Rossiter-McLaughlin results \citep{Petigura2015}. Thus we conclude that our measurement of $v\sin i_{\star} \sim 3\,\mathrm{km\,s}^{-1}$ is robust, while the previously reported value of $v\sin i_{\star} \sim 6\,\mathrm{km\,s}^{-1}$ represents an overestimate.

\subsection{Stellar Parameters}
We further estimated the stellar parameters using the procedure described by \citet{Fulton2018} and the \texttt{isoclassify} package \citep{Huber2017}. First, we applied the Stefan-Boltzmann law to derive the stellar radius ($R_{\star}$) based on the effective temperature from our Keck/HIRES spectroscopy, parallax measurements from Gaia~DR3 \citep{Gaia2021}, the $G$, $G_{\rm{BP}}$, and $G_{\rm{RP}}$ magnitudes from Gaia~DR3, and the $J$, $H$, and $K_{\rm{s}}$ magnitudes from the Two Micron All-Sky Survey. We then obtained various stellar parameters (e.g., stellar mass $M_{\star}$) from the posterior distributions by integrating over the MESA Isochrones \& Stellar Tracks \citep[MIST,][]{Choi2016} using the spectroscopic parameters, parallax, and their priors. As a result, we found $R_{\star} = 1.03 \pm 0.02\,R_{\odot}$ and $M_{\star} = 1.07 \pm 0.02\,M_{\odot}$, values that are consistent with $R_{\star} = 1.043 \pm 0.009\,R_{\odot}$ and $M_{\star} = 1.109 \pm 0.043\,M_{\odot}$ from \citet{Serrano2022}, as well as $R_{\star} = 1.03 \pm 0.02\,R_{\odot}$ and $M_{\star} = 1.08 \pm 0.03\,M_{\odot}$ from \citet{MacDougall2023}. We also derived a stellar age of $1.6_{-0.8}^{+1.0}$\,Gyr, which is consistent with the $1.4_{-0.4}^{+0.8}$\,Gyr reported in previous studies, although age determinations from isochrones generally have large uncertainties. Key stellar parameters are listed in Table~\ref{tab:stellar_para}.

\subsection{Stellar Rotation}\label{sec:prot_param}
We measured the stellar rotational period of HD~93963~A using \textit{TESS} photometry. HD~93963~A was observed by \textit{TESS} \citep{Ricker2014} during Sectors~22 and~48 at a 2-minute cadence, as shown in Figure~\ref{fig:tess_lc}. We downloaded the light curves from the Mikulski Archive for Space Telescopes \citep{MAST2021}\footnote{\url{https://archive.stsci.edu}}. In both Sectors~22 and~48, Figure~\ref{fig:tess_lc} reveals clear signs of quasi-periodic rotational modulation due to active regions on this early G-type star.

To determine the periodicity, we first performed a Lomb-Scargle analysis on each light curve separately. We identified the strongest peak at $\sim$$6.3$\,days in Sector~22, consistent with \citet{Serrano2022}, but a different period of $\sim$$4.1$\,days in Sector~48. Both periods had false alarm probabilities \citep[FAP;][]{Baluev2008} below $10^{-6}$. We then computed the auto-correlation function (ACF) of each light curve. As shown in the bottom panel of Figure~\ref{fig:tess_lc}, both ACFs exhibited correlated peaks at around $\sim 12.4$\,days, likely linked to stellar rotation. The ACF of the Sector~22 light curve has a second correlated peak at $12.8 \pm 1.6$\,days, while that of Sector~48 has a third peak at $12.1 \pm 1.3$\,days. Taking the uncertainty in the peak location to be the half-width at half-maximum, the two measurements differ by $0.3\,\sigma$. By combining results from Sectors~22 and~48, we obtained a weighted mean of $12.4 \pm 1.4$\,days. These findings also agree with radial velocity analyses using Gaussian Processes to within $1\,\sigma$ \citep{Serrano2022}. The photometric variation in periodicity may arise from changes in the number and distribution of active regions.

Using $P_{\rm{rot}} = 12.4 \pm 1.4$\,days, we derived a stellar age of $1.9_{-0.4}^{+0.5}$\,Gyr via gyrochronology \citep[\texttt{gyro-interp} package;][]{Bouma2023}, which is consistent with the isochrone age of $1.6_{-0.8}^{+1.0}$\,Gyr. Additionally, assuming a stellar inclination of $90^\circ$, our spectroscopic measurement of $v \sin i_{\star} = 2.93 \pm 1.00\,\mathrm{km\,s}^{-1}$ suggests a rotational period of $18_{-5}^{+9}$\,days, approximately 1.5 times of 12.4-day period from light curve. We could not identify a peak in this period range in either the GLS periodogram or the ACF, as it exceeds the Nyquist period of 13.5 days. However, we cannot fully rule out the possibility that the 12.4-day period is a harmonic of an 18-day period. Thus ensuring a sufficiently long baseline is critical to the measurement, yet the \textit{TESS} light curve still remains the primary limitation.

\section{Joint Light Curve and RM Analysis} \label{sec:rm}
\begin{deluxetable}{lcc}
\tablecaption{Stellar and Transit Parameters of HD~93963~Ac} \label{tab:planet_para}
\tablehead{
\colhead{Parameters} & \colhead{Posteriors}  & \colhead{Priors}}
\startdata
$\ln \rho$ ($\rho_{\odot}$)  & $-0.087_{-0.067}^{+0.068}$ & $\mathcal{N}[-0.9, 0.1]$ \\
$P_{\rm orb}$ (d)   & $3.6451392 \pm 0.0000037$ & $\mathcal{U}[3.635, 3.655]$ \\
$t_0$ (BJD-2458902) & $0.87377_{-0.00059}^{+0.00061}$ & $\mathcal{U}[0.77, 0.97]$ \\
$R_{\rm{p}}/R_{\star}$  & $0.0287 \pm 0.0003$ & $\mathcal{U}[0.005, 0.05]$  \\
$\cos\,i_{\rm{orb}}$ & $0.063 \pm 0.004$ & $\mathcal{U}[0, 1]$ \\
$\sqrt{v\sin i_{\star}}\cos\lambda$  & $1.61_{-0.23}^{+0.17}$& $\mathcal{U}[-10, 10]$ \\
$\sqrt{v\sin i_{\star}}\sin\lambda$  & $0.41_{-0.56}^{+0.50}$ & $\mathcal{U}[-10, 10]$ \\
$\gamma$ ($\rm{m}\,\rm{s}^{-1}$) & $-2.02 \pm 0.46$ & $\mathcal{U}[-1000, 1000]$ \\
$\dot{\gamma}$ ($\rm{m}\,\rm{s}^{-1}\,\rm{d}^{-1}$) & $-36.64_{-15.70}^{+15.57}$ & $\mathcal{U}[-1000, 1000]$ \\
$\ddot{\gamma}$ ($\rm{m}\,\rm{s}^{-1}\,\rm{d}^{-2}$) & $113.08_{-75.43}^{+76.50}$ & $\mathcal{U}[-1000, 1000]$ \\
$u_{1,\rm{TESS}}$& $0.2543_{-0.1034}^{+0.1025}$ & $\mathcal{N}[0.25, 0.14]$\\
$u_{2,\rm{TESS}}$& $0.2655_{-0.1147}^{+0.1173}$ & $\mathcal{N}[0.26, 0.14]$ \\
$u_{1,\rm{RM}}$& 0.4270 & Fixed \\
$u_{2,\rm{RM}}$& 0.2799 & Fixed \\
$\ln S_{0}$ ($\rm{m}^{2}\,\rm{s}^{-2}\,\mu\rm{Hz}^{-1}$)  & $-7.10_{-1.13}^{+0.95}$ & $\mathcal{U}[-10, 10]$ \\
$\ln \omega_{0}$ ($\mu\rm{Hz}$) & $5.72_{-0.53}^{+0.50}$ & $\mathcal{U}[4.3, 6.6]$ \\
$\ln Q$  & $\ln (1/\sqrt{2})$ & Fixed \\
$\ln \sigma$ ($\rm{m}\,\rm{s}^{-1}$)  & $-0.73_{-0.39}^{+0.30}$ & $\mathcal{U}[\ln(0.25), \ln(10)]$ \\
\hline
$\lambda$ ($^{\circ}$)  & $14_{-19}^{+17}$ & Derived \\
$v\sin i_{\star}$ ($\rm{km}\,\rm{s}^{-1}$)  & $2.96 \pm 0.64$ & Derived$^{\ddagger}$ \\
$R_{\rm p}$ ($R_\oplus$)  & $3.230 \pm 0.072$ & Derived\\
$T_{14}$ (hrs) &  $2.876_{-0.064}^{+0.065}$ & Derived \\  
$b$   & $0.616_{-0.025}^{+0.023}$& Derived \\
$i_{\rm{orb}}$ ($^{\circ}$)  & $86.35_{-0.21}^{+0.22}$& Derived\\ 
$a/R_\star$ &  $9.68_{-0.21}^{+0.22}$& Derived  \\
$\sigma$ ($\rm{m}\,\rm{s}^{-1}$) & $0.48_{-0.16}^{+0.17}$ & Derived\\
\enddata
\tablecomments{In the joint model, the eccentricity is fixed to zero ($e=0$). $^{\ddagger}$This $v\sin i_{\star}$ result is derived from RM effect.}
\end{deluxetable}

Two transiting planets (b and c) orbiting HD~93963~A were confirmed from the \textit{TESS} light curve. To obtain clear transit signals, we detrended the stellar variability in the Sectors 22 and 48 light curves. Since our focus was on planet c, we masked all transits of planet b. We determined the sky-projected obliquity by combining the in-transit light curve, the RM effect, and the RV noise and trend. Prior to modeling, we segmented the light curves into individual transit events to facilitate the analysis.

\subsection{Light Curve and RM Modeling}
For light-curve modeling, we used the \texttt{batman} package \citep{Kreidberg2015} and adopted a quadratic limb-darkening law, allowing its coefficients ($u_{1,\rm{TESS}}$, $u_{2,\rm{TESS}}$) to vary as free parameters. Following \citet{Serrano2022}, we fixed both eccentricity and argument of periastron to zero. Other free parameters included the logarithm of stellar density $\ln\rho_{\star}$ with a Gaussian prior, the orbital period $P_{\rm{orb}}$, the planet/star radius ratio $R_{\rm{p}}/R_{\star}$, the time of conjunction $t_0$ and cosine of orbital inclination $\cos i$. We set wide, uniform priors for all parameters; specifically, $\cos i$  ranged from 0 to 1 and was converted into the impact parameter $b$.

For modeling the RM effect, we followed the approach of \citet{Hirano2011}, which accounts for stellar rotation, macroturbulence, thermal and pressure broadening, and instrumental broadening. We set the macroturbulence to $4.1\,\rm{km}\,\rm{s}^{-1}$ \citep{Serrano2022}, consistent with the $4.3\,\rm{km}\,\rm{s}^{-1}$ predicted by the \citet{Valenti2005} scaling relation. Thermal and instrumental broadening were calculated following \citet{Hirano2011}, while pressure broadening was fixed at $1\,\rm{km}\,\rm{s}^{-1}$ for a G-type star. We also used a quadratic limb-darkening law with coefficients ($u_{1,\rm{RM}}$, $u_{2,\rm{RM}}$) set to the values derived by \texttt{EXOFAST}\footnote{\url{https://astroutils.astronomy.ohio-state.edu/exofast/limbdark.shtml}} for the stellar parameters ($T_{\rm{eff}}$, $\rm{[Fe/H]}$, $\log g$) in Section \ref{sec:star}. Free parameters included the sky-projected obliquity $\lambda$ and the projected stellar rotational velocity $v\sin i_{\star}$. For improved sampling efficiency, these were transformed into $\sqrt{v\sin i_{\star}}\cos\lambda$ and $\sqrt{v\sin i_{\star}}\sin\lambda$, with uniform priors assigned to both.

\subsection{Noise and Trend Modeling}
The amplitude of the RM effect for HD~93963~Ac ($\sim$1 m s$^{-1}$) is comparable to the asteroseismic variability, so the seismic jitter in the RV time series cannot necessarily be ignored or simply modeled as white noise. According to the scaling relation presented by \citet{Kjeldsen1995}, the p-mode oscillation timescale of the star is $5.5\pm0.2$ minutes. Because this is shorter than the 6-min exposure time, the effect of $p$-modes on the measured radial velocities should be minimal. According to scaling relation in \citet{Chaplin2019}, the RV residual cased by $p$-modes can be suppressed down to $\sim$0.3 m s$^{-1}$. Meanwhile, the scaling relation in \citet{Kallinger2014} indicates a granulation timescale of $22\pm1$ minutes, which may persist as correlated noise in the time series.

To account for the correlated noise caused by granulation, we added a Gaussian Process (GP) in our joint modeling. We adopted a stochastically-driven damped simple harmonic oscillator (SHO) kernel, whose power spectrum density can be expressed as:
\begin{equation}
    S(\omega) = \sqrt{\frac{2}{\pi}}\frac{S_{0} \omega_{0}^{4}}{(\omega^{2} - \omega_{0}^{2})^{2} + \omega_{0}^{2}\omega^{2}/Q^{2}},
\end{equation}
where $\omega_{0}$ is the frequency of the undamped oscillator, $S_0$ is proportional to the power at $\omega=\omega_{0}$, and $Q$ is the quality factor of the oscillator. Following \citet{Foreman-Mackey2013}, we fixed the quality factor $Q=1/\sqrt{2}$ for granulation modeling. Then the GP kernel can be expressed as:
\begin{equation}
    k(\tau) = S_{0}\omega_{0} e^{-\frac{1}{2}\omega_{0}\tau}\cos\left (\frac{\omega_{0}\tau}{\sqrt{2}}-\frac{\pi}{4}\right),
\end{equation}
where $\tau$ is the time difference of two measurements. Therefore, there were two free parameters, $S_{0}$ and $\omega_{0}$, and we used their logarithm form, $\ln S_{0}$ and $\ln \omega_{0}$, in the GP model. They were set with uniform priors. For $\omega_0$, a lower limit was imposed based on half of the observational timespan, and an upper limit was imposed based on twice the time sampling (12 minutes).

To account for additional noise sources, we included an extra jitter term in logarithmic form, $\ln\sigma_{\rm{jit}}$, with a uniform prior in the model. Notably, convective blueshift can be absorbed into this jitter term rather than modeled separately. We estimated the convective blueshift following the formulation in \citet{Shporer2011}, finding that the resulting RV amplitude could be less than $0.2\, \rm{m}\,\rm{s}^{-1}$ which is weaker than the typical observational uncertainty of each RV measurement.

Additionally, we modeled the intra-night RV variation caused by long-term stellar activity and planetary orbital motion using a second-order polynomial, consisting of the RV zero-point offset $\gamma$, linear RV trend $\dot{\gamma}$, and quadratic RV trend $\ddot{\gamma}$. Uniform priors were also applied to these parameters.

\subsection{MCMC Sampling}

We sampled the posterior distribution of our joint RM+Transit+GP model using a Markov Chain Monte Carlo (MCMC) algorithm, as implemented in the \texttt{emcee} code \citep{Foreman-Mackey2013}. 
We generated MCMC chains with 32 random walkers and the priors in Table \ref{tab:planet_para} and initialized their positions around perturbed best-fit models. These best-fit values were obtained by maximizing the likelihood function via the \texttt{Powell} algorithm \citep{Powell1964}, as implemented in the \texttt{scipy.optimize} package \citep{SciPy-NMeth2020}. To finalize the results, we continued the long-chain MCMC until it satisfied the Gelman-Rubin convergence criterion with $\rm{GR} < 1.01$.

\subsection{Fitting Results}\label{sec:rm_res}
Consequently, we obtained a sky-projected obliquity of $\lambda = 14_{-19}^{+17\circ}$ from the posterior, favoring a prograde and aligned orbit for HD~93963~Ac. We also derived a projected stellar rotational velocity of $v \sin i_{\star} = 2.96_{-0.64}^{+0.65}\,\mathrm{km\,s}^{-1}$, which well matches the spectroscopic result from the Keck/HIRES spectra and suggests that the previously reported value of $v \sin i_{\star} = 5.9 \pm 0.8\,\mathrm{km\,s}^{-1}$ was overestimated. The granulation timescale, converted from the oscillation frequency $\omega$, was estimated to be $30_{-12}^{+21}$ minutes, which appears slightly overestimated at the 1-$\sigma$ level compared to the theoretical value. In addition, we did not find evidence of overfitting in our model, as indicated by a reduced $\chi_{\rm{red}}^{2}$ of 1.01. All fitted and derived parameters from our modeling are listed in Table \ref{tab:planet_para}.

\subsection{HD~93963 system in 3D}
The 3D architecture, including the stellar inclination and true obliquity, can be constrained by combining the known stellar rotation and projected stellar velocity, following the method suggested by \citet{Masuda2020}. We adopted the best-constrained rotational period of $P_{\rm{rot}} = 12.4 \pm 1.4\,\mathrm{d}$ from this work and a projected rotational velocity of $v \sin i_{\star}=2.96\,\mathrm{km\,s}^{-1}$ from our RM-effect fit, while conservatively assigning an uncertainty of $2\,\mathrm{km\,s}^{-1}$. The true obliquity $\psi$ can be expressed as \citep{Fabrycky2009}:
\begin{equation}
    \cos \psi = \cos i_{\rm{orb}} \cos i_{\rm{\star}} + \sin i_{\rm{orb}} \sin i_{\rm{\star}} \cos \lambda,
\end{equation}
where $i_{\rm{\star}}$ is the stellar inclination, $i_{\rm{orb}}$ is the orbital inclination, and $\lambda$ is the sky-projected obliquity. Given the substantial uncertainty in the $v \sin i_{\star}$ and stellar rotation period measurements, as addressed in Section \ref{sec:atm_param} and \ref{sec:prot_param}, we can only place upper limits on the stellar inclination and true obliquity: 
$i_{\star} < 34^\circ$ at 84\% confidence and $\psi < 56^\circ$ at 84\% confidence. Or, assuming $P_{\rm{rot}} = 18 \pm 5\,\mathrm{d}$, we can obtain $i_{\star} > 64^\circ$ at 84\% and $\psi < 53^\circ$ at 84\% confidence.

The HD~93963 system likely hosts a massive companion in a wide orbit. \citet{Serrano2022} reported a long-term trend in their RV time series. \citet{VanZandt2025} assessed a planet with $P = 870 \pm 36$ days and $M\sin\,i = 0.9 \pm 0.1\,M_{\mathrm{J}}$. Considering that the star is at a distance of 82.3~pc away from us and has a Gaia RUWE value of 1.0, the outer planet candidate lies beyond Gaia's current detection limit. Gaia DR4 may provide an opportunity to constrain its orbital inclination.

\section{Transmission Spectroscopy}
\begin{figure*}
    \centering
    \includegraphics[width=.8\linewidth]{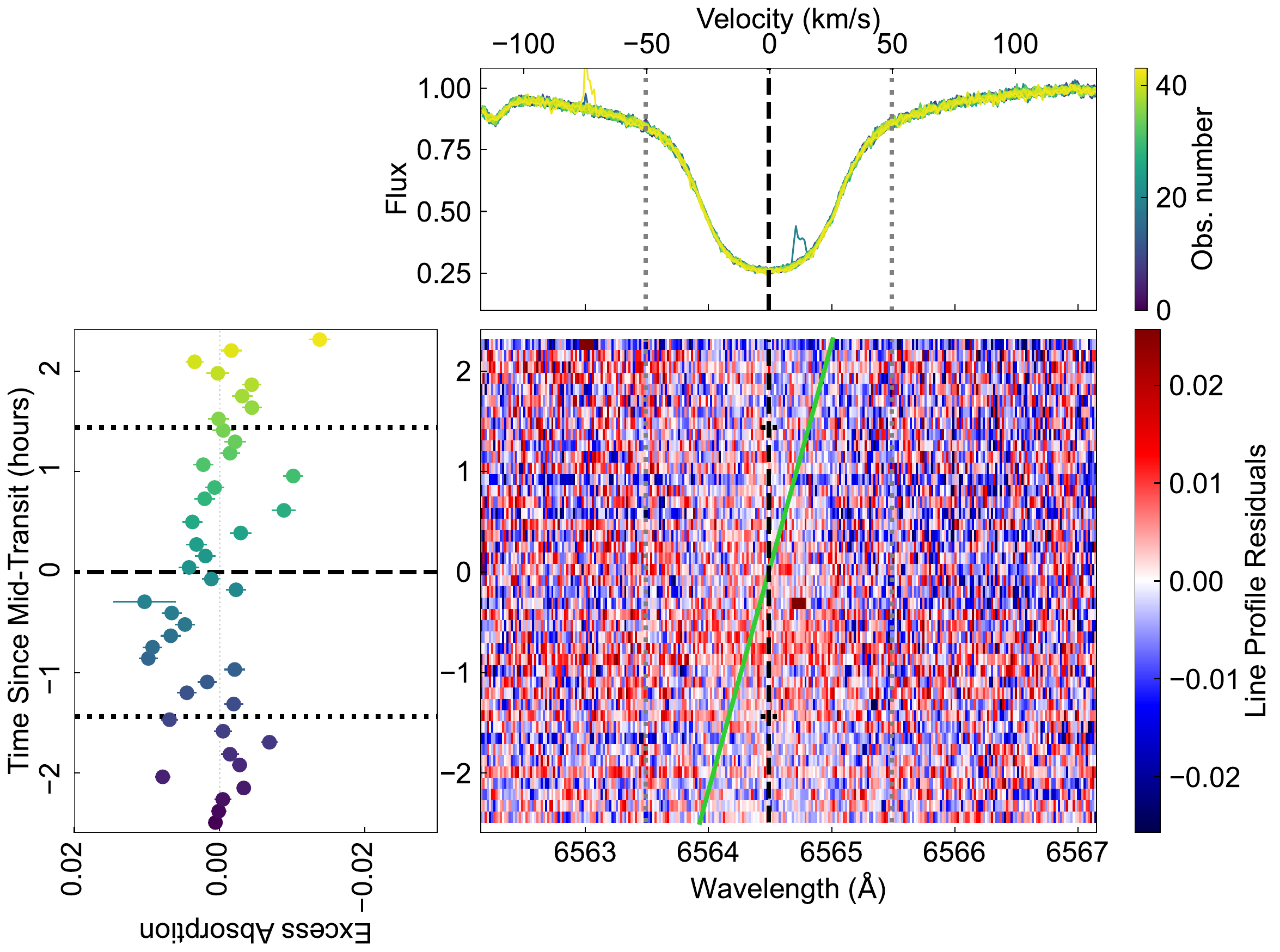}
    \caption{Top: Normalized spectra of H$\alpha$ line \AA\, during transit event of planet c with their color indicating their observation number (epoch). The dashed line marks the line center at 6564.7 \AA, and the dotted lines indicate 1 \AA\, from the line center. Left: Excess absorption of H$\alpha$ variation by time, with the absorption referenced by a spectrum averaged from pre-ingress and post-egress spectra, and calculated in a bandpass of 2 \AA\, centered at 6564.7 \AA. The color scale is the same for the top sub-plot. The dashed line and dotted line indicate mid-transit and ingress/egress, respectively. Main (Bottom right): The measured H$\alpha$ line profile residuals as a function of time and wavelength. The dashed line and vertical dotted line correspond to the top sub-plot the top subplot, while the horizontal dotted line indicates the egress as in the left subplot. The predicted planetary velocity is marked by green solid line. Neither excess absorption nor color shift occurred during the transit event.}
    \label{fig:halpha}
\end{figure*}

The low density of HD~93963~Ac, $\rho = 3.1\,\rm{g\,cm}^{-3}$, suggests that the planet is not completely solid and likely retains an atmosphere. At an age of $1.6_{-0.8}^{+1.0}$ Gyr, photoevaporation should be significantly slower than it was in the first few Myr after formation. Nevertheless, to check for any evidence of mass loss, we examined specific lines within Keck/KPF's wavelength coverage, including \ion{Mg}{1} at 5185\,\AA, H$\alpha$ at 6564\,\AA, \ion{Na}{1} at 5892\,\AA, and \ion{Ca}{2} at 8544\,\AA, that may indicate ongoing atmospheric loss \citep{Linssen2023}. An example of the H$\alpha$ line is shown in Figure~\ref{fig:halpha}. As expected, we did not detect any excess absorption in the transmission spectra, quantified by the equivalent width of the line profile residuals (left panel of Figure~\ref{fig:halpha}). Our null detection implies that, by this age, the planet may have already completed most of its photoevaporation. \citet{Serrano2022} estimated a mass-loss timescale of about 150~Myr.

\section{Dynamics of HD~93963 system}\label{sec:dyn}
In this section, we present both a secular and an N-body analysis of the future evolution of HD~93963 given the observed orbital architecture.

\subsection{Analytical theory of nodal precession}\label{sec:ana_sol}
Since HD~93963~Ab and Ac are far from resonance and likely have a small true mutual inclination \citep{Serrano2022}, we used the Laplace-Lagrange secular theory to examine their dynamical evolution \citep{Murray1999}. We made two additional assumptions: (1) the planets are in circular orbits and have a small mutual inclination, and (2) the stellar spin direction is unvarying, i.e., 
the precession of the stellar orientation due to planetary orbital angular momenta is neglected
(following \citealt{Spalding2016}, and justified below). We aligned the $z$ axis with the stellar rotation axis. The true stellar obliquity $\psi$ thereby set the initial inclination of the planetary orbits relative to the $x$--$y$ plane.

Within this framework, the dynamical evolution of the inclination of the planetary orbits, and the ascending node, $\Omega$, can be expressed as a linear differential equation,
\begin{equation}\label{eqn:xi}
    \frac{d\boldsymbol{\xi}}{dt} = -\iota\, \boldsymbol{M}  \cdot \boldsymbol{\xi},
\end{equation}
where $\boldsymbol{\xi}$ is the complex inclination vector with $\xi_{i} = \psi_{i}\cos\Omega_i + \iota \psi_{i}\sin\Omega_i$, $\iota$ is imaginary unit, and the matrix $\boldsymbol{M}$ describes the nodal precession due secular effects and additional effects,
\begin{equation}
\begin{aligned}
    M_{ii} &= \nu_{i} + \sum^{N}_{j=1, j\neq i} B_{ij} \\ 
    M_{ij} &= - B_{ij},
\end{aligned}
\end{equation}
where $B_{ij}$ is the precession rate induced on planet $i$ by planet $j$ and $\nu_{i}$ is the sum of precession rate induced on planet $i$ by additional effects. For each planet, the precession rate by another planet can be described as,
\begin{equation}
    B_{ij} \equiv n_{i}  \frac{m_{j}}{4 M_{\star}} b_{3/2}^{(1)} \alpha_{ij} \Bar{\alpha}_{ij}, \\
\end{equation}
where $\alpha_{ij} = a_{i}/a_{j}$, $\bar{\alpha}_{ij} = 1$ if $a_{i} > a_{j}$ and otherwise $\Bar{\alpha}_{ij} = \alpha_{ij} = a_{i}/a_{j}$. In the equations above, $a_i$ and $n_i=\sqrt{G(M_{\star}+m_i)/a_i^3}$ are respectively the semimajor axis and mean motion of the planet $i$, and $b_{3/2}^{(1)}$ is the Laplace coefficient.
In this framework, we only consider the host star's leading order quadrupole effect on the planet. Other effects, such as higher order general relativity, can be negligible. Hence for each planet, the precession rate can be described as,
\begin{equation}
    \nu_i \equiv \nu_{\star,i}.
\end{equation}
The precession induced by host star's quadrupole can be described as,
\begin{equation}
    \nu_{\star,i} = \frac{3}{2} n_{i} J_{2} \left( \frac{R_{\star}}{a_i}\right)^{2},
\end{equation}
where the quadrupole moment $J_2$ can be calculated through the following approximate relationship \citep{Spalding2020},
\begin{equation}
    J_{2} \approx 10^{-3} \left(\frac{k_{2}}{0.2}\right) \left(\frac{P_{\rm{rot}}}{\rm{day}}\right)^{-2} \left(\frac{R_{\star}}{R_{\odot}}\right)^{3} \left(\frac{M_{\star}}{M_{\odot}}\right)^{-1},
\end{equation}
with $k_2\approx0.2$ assuming the star is fully convective according to \citet{Batygin2013}.

The general solution of the differential equation (Equation \ref{eqn:xi}) is the sum of normal modes, and it can be written in terms of eigenvalues, $\lambda_{k}$, and eigenvectors, $\boldsymbol{\xi}_{k}$, as:
\begin{equation}\label{eqn:sol}
    \boldsymbol{\xi}(t) = \sum_{k}^{} c_{k}  \exp(-\iota \lambda_{k} t) \boldsymbol{\xi}_{k},
\end{equation}
where the constants, $c_{k}$, can be determined from the initial condition. The constants can also be assembled into a vector, $\boldsymbol{c}$, and expressed as:
\begin{equation}
\boldsymbol{c} = (\boldsymbol{X}^{\rm{T}}\cdot\boldsymbol{X})^{-1}\cdot\boldsymbol{X}^{\rm{T}}\cdot\boldsymbol{\xi}(t=0),
\end{equation}
where $\boldsymbol{X}$ is a matrix consisting of the eigenvectors:
\begin{equation}
\boldsymbol{X} = \left( \boldsymbol{\xi}_{1}, \boldsymbol{\xi}_{2}, ..., \boldsymbol{\xi}_{n} \right).
\end{equation}
A more detailed derivation of the equations above can be found in \citet{Murray1999}. 

\subsection{Testing on HD~93963 system}\label{sec:dyn_ana_93963}
\begin{deluxetable*}{cccccccc}[ht]
\tablecaption{The necessary planetary parameters in Equation \ref{eqn:xi} and calculated precession rates of three planets induced by planets and stellar quadrupole in Equation \ref{eqn:xi}. } \label{tab:precession_rate}
\tablehead{
Planet & $a\, (\rm{AU})$ & $m\, (\rm{M_{\oplus}})$   & $B_{\rm{b}\it{j}}\, (\rm{rad}\, \rm{yr}^{-1})$ & $B_{\rm{c}\it{j}} \, (\rm{rad}\, \rm{yr}^{-1})$ & $B_{\rm{d}\it{j}}\, (\rm{rad}\, \rm{yr}^{-1})$ & $\sum{B_{ij}}\, (\rm{rad}\, \rm{yr}^{-1})$ &  $\nu_{i}\, (\rm{rad}\, \rm{yr}^{-1})$ } 
\startdata
b & $0.02085^{\rm{A}}$ & $4.4^{\rm{C}}$  & -- & -$9.63 \times 10^{-3}$ & $-1.64 \times 10^{-6}$ & $9.63 \times 10^{-3}$ & $1.81 \times 10^{-3}$ \\
c & $0.04813^{\rm{A}}$ & $18.4^{\rm{B}}$ & $-1.55 \times 10^{-3}$& -- & $-5.75 \times 10^{-6}$& $1.56 \times 10^{-3}$ & $9.69 \times 10^{-5}$  \\
d & $2.0^{\rm{D}}$ & $317.8^{\rm{D}}$   & $-2.31 \times 10^{-9}$ & $-5.05 \times 10^{-8}$& -- &  $5.28 \times 10^{-8}$ & $2.10 \times 10^{-10} $\\
\enddata
\tablecomments{Stellar quadrupole was set by $J_2 = 1\times10^{-5}$; A: Semimajor axes of planet b and c were both obtained from \citet{Serrano2022} for consistency; B: The mass of planet c was obtained from \citet{Polanski2024}; C: The mass of planet b was the weighted mean of the best-fit result from \citet{Serrano2022}, \citet{Polanski2024} and \citet{Brinkman2024}; D: We set planet d with a reasonable assumption.}
\end{deluxetable*}

\begin{deluxetable*}{ccccc}[ht]
\tablecaption{The eigensystem of the precession matrix $\boldsymbol{M}$ in Equation \ref{eqn:sol} specified by the parameters of Table \ref{sec:dyn_ana_93963}. } \label{tab:eigen}
\tablehead{
Eigenmode & No. 1 & No. 2 (dominant) & No. 3 & Planet }
\startdata
Eigenvalues $(\rm{rad}\,\rm{yr}^{-1})$ & $1.3 \times 10^{-2}$ & $3.1 \times 10^{-4}$ & $5.2 \times 10^{-8}$ & \\
\hline
  & $9.9 \times 10^{-1}$ & $-6.5 \times 10^{-1}$ & $1.5 \times 10^{-2}$ & b \\
Eigenvectors  & $-1.4 \times 10^{-1}$ & $-7.6 \times 10^{-1}$ & $1.7 \times 10^{-2}$ & c\\
& $3.7 \times 10^{-7}$ & $1.3 \times 10^{-4}$ & $ 1.0 $ & d \\
\hline
Constant $c_{k}$ & $6.5 \times 10^{-19} + 1.1\,i \times 10^{-2}$ & $-7.1 \times 10^{-18} -1.2\,i \times 10^{-1}$ & $5.3 \times 10^{-18} + 8.8\,i \times 10^{-2}$ &\\
\enddata
\tablecomments{Stellar parameters were adopted by the values derived in this work. Planetary parameters were adopted from Table \ref{tab:precession_rate}, and additionally, true stellar obliquities were set to $\psi_{\rm{b}} = \psi_{\rm{c}} = \psi_{\rm{d}}= 5^{\circ}$ and ascending nodes were set to $\Omega_{\rm{b}} = \Omega_{\rm{c}} = \Omega_{\rm{d}} = 90^{\circ}$. }
\end{deluxetable*}

\begin{figure*}[ht]
\centering
    \includegraphics[width=0.495\linewidth]{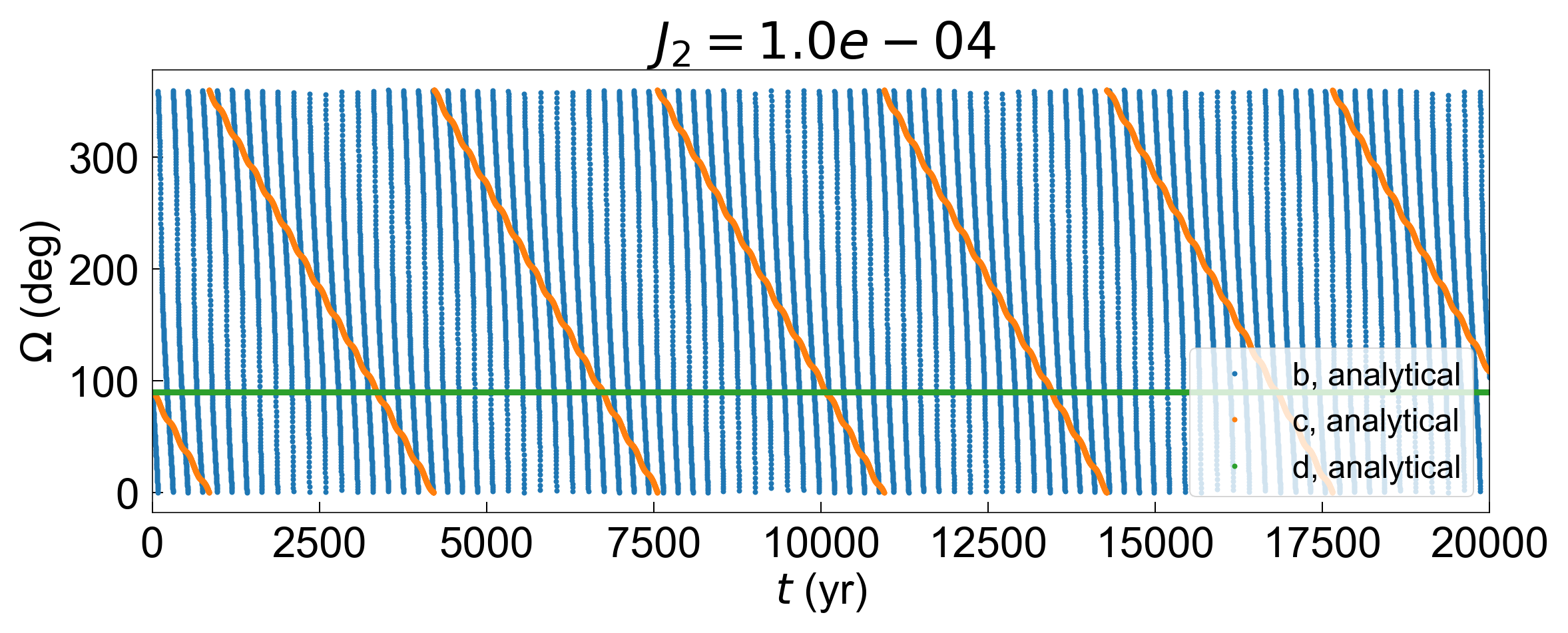}
    \includegraphics[width=0.495\linewidth]{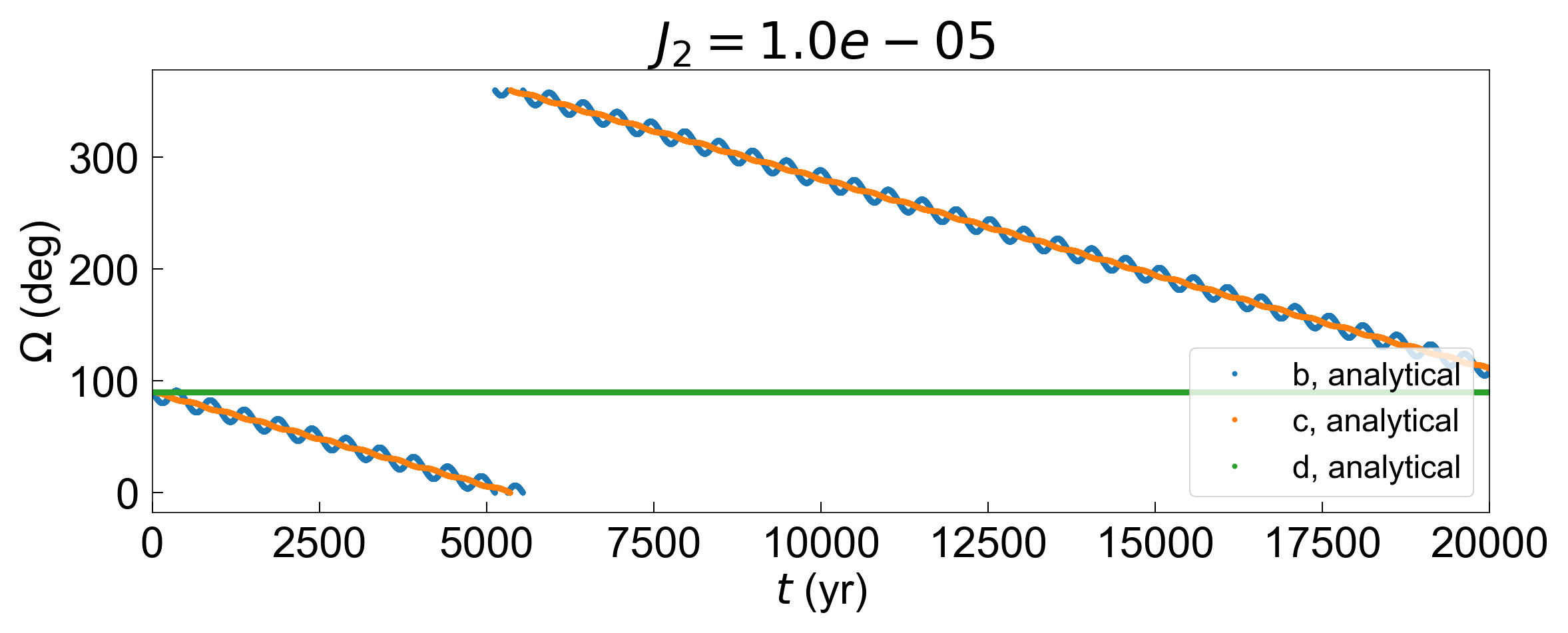}
    \caption{The dominant precession pattern of the planet b (USP) changes as $J_2$ value changes. The left panel shows the star having $J_2 = 1.0\times10^{-4}$, where planet b and c precess follow different eigenmodes, while the right panels shows the stars having $J_2 = 1.0\times10^{-5}$ where planet b and c mainly precess at a same eigenmode. A comparison providing various $J_2$ values is in Figure \ref{fig:J2_variation_appendix} in Appendix.}
    \label{fig:J2_variation}
\end{figure*}

We modeled the HD~93963 system using Laplace-Lagrange theory as outlined in the previous section. We assumed that the system consists of its confirmed planets, b and c (inner two planets), as well as an outer giant planet companion on a nearly coplanar orbit sugguested by long-term RV trend \citep{VanZandt2025}, to examine the secular effects that the outer planet might induce on the two inner planets. 

For the stellar parameters, we adopted the mass and radius values from Table \ref{tab:stellar_para} and the stellar rotational period derived in this work. Using these parameters, we estimated a $J_2$ value of $1 \times 10^{-5}$. For the planetary parameters, we used the masses and semimajor axes listed in Table \ref{tab:precession_rate}. Specifically, the outer giant planet companion was assumed to have a mass of $1 M_{\rm{J}} = 317.8\, M_{\rm{\oplus}}$ and orbit the star at 2 AU, based on \citet{Serrano2022} and \citet{VanZandt2025}.  We estimated that the timescale of stellar spin precession induced by this outer giant planet companion could exceed 10 Gyr \citep[Equation 19]{Storch2015}. Furthermore, the stellar companion HD~93963 B ($170\, M_{\rm{J}}$, 484 AU away) would induce a precession on the stellar spin at a timescale of $10^4$ Gyr. Consequently, the stellar orientation can be considered fixed, consistent with the second assumption in our framework. 

Using the stellar and planetary parameters outlined above, we calculated the nodal precession rates induced on each planet using Equation \ref{eqn:xi}. The key results are summarized in Table \ref{tab:precession_rate}, with two notable implications. First, the nodal precession rates of the two inner planets induced by the outer giant planet companion (approximately $10^{-8}\, \rm{rad}\, \rm{yr}^{-1}$) are significantly lower than the rates they induce on each other (approximately $10^{-3}\, \rm{rad}\, \rm{yr}^{-1}$). Furthermore, the nodal precession induced by the distant stellar companion occurs on a timescale of $10^3$ Gyr. These indicate that, over shorter timescales (e.g., $\sim$10 Myr), the outer companion's influence on the inner planets' nodal precession can be neglected. Second, the inner planets share precession rates on the order of $10^{-3}\, \rm{rad}\, \rm{yr}^{-1}$, suggesting that planets b and c are dynamically coupled into coplanarity.

Our framework enables evaluating the coplanarity of the inner two planets in HD~93963 system by identifying their dominant eigenmode, i.e., the dominant precession period. Using the matrix elements in Table \ref{tab:precession_rate}, we calculated three eigenvalues and their corresponding eigenvectors, listed in descending order in Table \ref{tab:eigen}. Determining the dominant eigenmode requires initial conditions to compute the constant factors in Equation \ref{eqn:xi}. For illustration, we assume true stellar obliquities of $\psi_{\rm{b}} = \psi_{\rm{c}} = \psi_{\rm{d}} = 5^{\circ}$ and ascending nodes of $\Omega_{\rm{b}} = \Omega_{\rm{c}} = \Omega_{\rm{d}} = 90^{\circ}$. The resulting constant factors are provided in Table \ref{tab:eigen}.

We first identified that the lowest eigenvalue, $\lambda_{3} = 5.2 \times 10^{-8}\,\rm{rad}\,\rm{yr}^{-1}$, in Table \ref{tab:eigen} corresponds to the nodal precession rates of the two inner planets induced by the outer giant planet companion. This eigenmode can be neglected when analyzing the system on short timescales. We then focused on the highest and intermediate eigenmodes. For the intermediate eigenmode ($\lambda_{2} = 3.2 \times 10^{-4}\,\rm{rad}\,\rm{yr}^{-1}$), the product of the eigenvector and constant factor for planets b and c is roughly an order of magnitude larger than for the highest eigenmode ($\lambda_{1} = 1.3 \times 10^{-2}\,\rm{rad}\,\rm{yr}^{-1}$). This indicates that planets b and c predominantly precess according to the second eigenmode, suggesting they are nearly coplanar. Additionally, this finding implies that nodal precession is mainly driven by secular interactions between the planets. While the stellar quadrupole accelerates precession rates across all eigenmodes, it does not  alter the dominant eigenmode for any planet.

The true stellar obliquities of $\psi_{\rm{b}} = \psi_{\rm{c}} = \psi_{\rm{d}} = 5^{\circ}$ were chosen purely for illustrative purposes, and the value of $5^{\circ}$ can represent any arbitrarily small angle. Similarly, the ascending node $\Omega$ can have an arbitrary but identical value for planets b and c, while for planet d, it does not need to match since the influence of planet d’s precession on planets b and c is negligible over short timescales. For this reason, adopting a perturbation model, as described by \citet{Lai2017} and \citet{Spalding2020}, is unnecessary for the HD~93963 system.

To determine when the stellar quadrupole significantly impacts nodal precession, we analyzed its dependence on the $J_{2}$ value. By varying $J_{2}$ from $1\times10^{-4}$ times to $1\times10^{-5}$, while holding other parameters unchanged, we calculated the ascending node's variation over time using Equation \ref{eqn:sol}. The results indicate that planets b and c precess independently on different eigenmodes when $J_2 \gtrsim 0.8 \times 10^{-5}$ but precess together in the same eigenmode, forming a coplanar pattern, when $J_2 \lesssim 0.3 \times 10^{-5}$. Figure \ref{fig:J2_variation} compares cases with $J_2 = 1 \times 10^{-4}$ and $J_2 = 1 \times 10^{-5}$. These findings suggest that when the host star HD~93963~A was young, with faster rotation and a stronger quadrupole effect, planets b and c likely precessed independently at different eigenmodes. As the star’s rotation slowed, the two planets may have transitioned into a coplanar configuration, but this alignment could deteriorate over time.

\subsection{N-body simulation and analytical solution}\label{sec:nbody-comp}
In the Laplace-Lagrange model, the evolution of inclination and the ascending node can be easily obtained by solving Equation \ref{eqn:sol}. For comparison, we performed N-body simulations to further verify our analytical results. The acceptable difference between these two solutions within a certain timescale can make the planets' behavior predicted more easily with lower computational cost. 
In our N-body simulation, we used the \texttt{REBOUND} code with the \texttt{WHFast} integration algorithm \citep{Rein2012,Rein2015} at a step size of $1/30$ of the orbital period of the innermost planet. We also used the \texttt{REBOUNDx} code \citep{Tamayo2020} to incorporate additional physical effects, including the $J_{2}$ effect (i.e., stellar quadrupole, using the \texttt{gravitational\_harmonics} module) and the general relativity modification (using the \texttt{gr} module). The stellar spin orientation was also set to align with the z-axis and be consistent with the analytical model.

We tested the HD~93963 system using identical initial orbital configurations for both the analytical solution and the N-body simulation. In each run, the initial true stellar obliquities of the three planets were set to the same value. The initial ascending nodes were assigned in two ways: (1) the inner two planets shared the same value, while the outer giant planet had an arbitrary value, ensuring a coplanar configuration for the inner planets; or (2) all three planets were given arbitrary values, leading to a non-coplanar configuration where the inner planets precess independently.

For both coplanar and non-coplanar cases, we computed the system's evolution over 20 kyr (about 40 times the dominant eigen period of $\sim$490 years) using both methods. We tested 100 initial true stellar obliquities ranging from $0.1^{\circ}$ to $15^{\circ}$, noting that the analytical solution is valid only for small inclination configurations \citep{Spalding2016}. For each initial obliquity value, 20 runs were performed with random initial ascending nodes. The results showed good agreement between the N-body simulation and the Laplace-Lagrange model over timescales of a few kyr, demonstrating consistency for the HD~93963 system.

\subsection{True obliquity and mutual inclination}
\begin{figure*}
    \centering
    \includegraphics[width=0.49\linewidth]{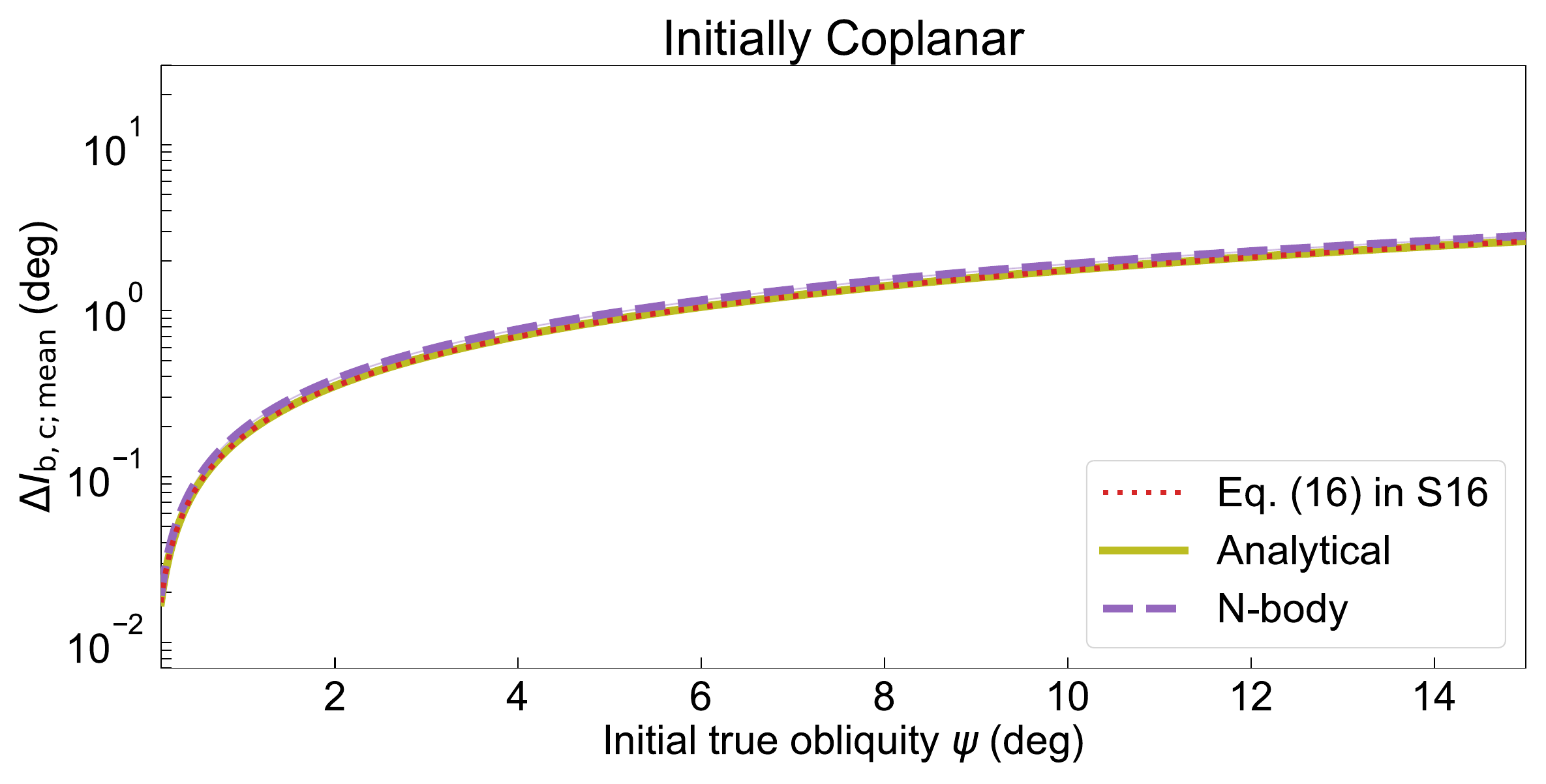}
    \includegraphics[width=0.49\linewidth]{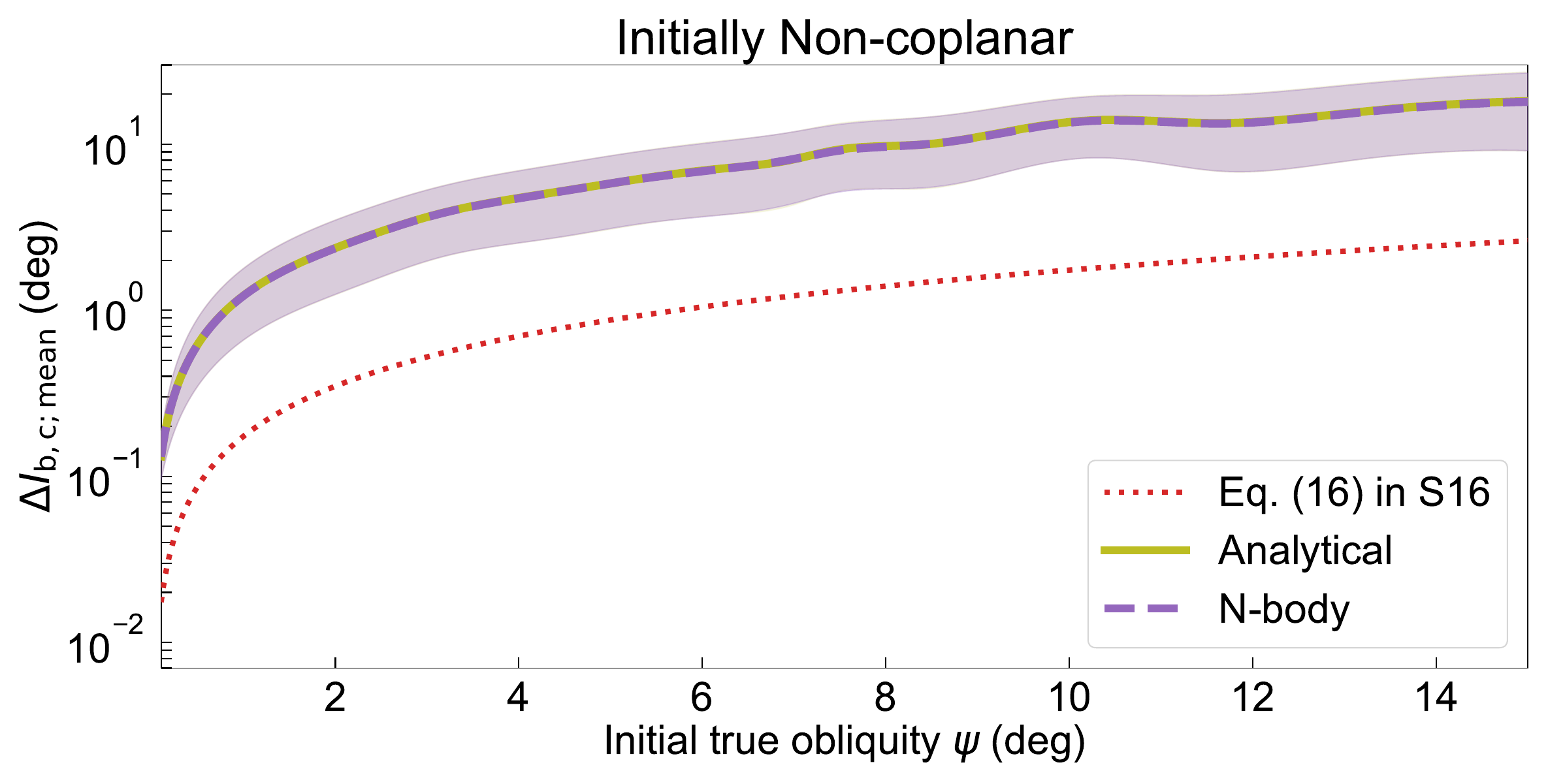}
    \caption{The estimated mutual inclination between HD~93963~Ab and Ac as a function of the initial true stellar obliquity ($\psi$). The left and right panels are respectively initialized from coplanar and non-coplanar configurations. The purple and olive colors represent results from the N-body simulation and the analytical solution, respectively. The red dotted lines mark the prediction by Equation 16 in \citet{Spalding2016} under the coplanar initialization.}
    \label{fig:mut_inc_inc}
\end{figure*}

In this subsection, we build on our dynamic analysis (Sections \ref{sec:ana_sol} and \ref{sec:nbody-comp}), where the variation in true stellar obliquity over time was determined using both the analytical solution and the N-body simulation. We then place a rough constraint on the true stellar obliquity by examining the true mutual inclination of the system’s planets.

First, we calculated the mutual inclination of the inner two planets in the HD~93963 system for each simulation run. The true mutual inclination between two planets is given by
\begin{equation}
    \cos \Delta I = \cos\psi_{1}\cos\psi_{2} + \sin\psi_{1}\sin\psi_{2}\cos(\Omega_{2}-\Omega_{1}).
\end{equation}
For each run, we calculated the time-averaged mutual inclination. Since multiple runs were conducted for each initial true stellar obliquity, with different ascending nodes, we derived an average mutual inclination for each true obliquity. Repeating this procedure across all true stellar obliquities allowed us to map out how the inner two planets’ true mutual inclination depends on the initial true stellar obliquity.

We tested both our analytical solution and the N-body simulation. Figure \ref{fig:mut_inc_inc} shows the mutual inclination for coplanar and non-coplanar initial configurations of HD~93963~Ab and Ac. The two methods yield nearly identical results, indicating that the analytical solution can serve as a computationally efficient substitute for the N-body simulation.

For the coplanar initialization, our results also match Equation (16) in \citet{Spalding2016}, which describes true mutual inclination evolving in a sinusoidal pattern if the system begins coplanar, on short timescales: 
\begin{equation}\label{eqn:s16_eq16}
    \Delta I(t) = 2 \psi \mathcal{G}(J_{2}) \sin (\omega_{0} t/2),
\end{equation} 
where the semi-amplitude $\mathcal{G}(J_{2})$ and frequency $\omega_{0}$ are derived in \citet{Spalding2016}. 

Notably, non-coplanar initialization yields larger true mutual inclinations for the same initial true stellar obliquity. For example, an initial true stellar obliquity of $5^\circ$ produces a time-averaged true mutual inclination of $\Delta I_{\rm{b,c}} = 1^\circ$ in the coplanar case, but $\Delta I_{\rm{b,c}} = 5^\circ$ when the system is initialized non-coplanar. As discussed in Section \ref{sec:dyn_ana_93963}, HD~93963~Ab and Ac share the same dominant eigenmode under the current stellar quadrupole effect ($J_{2} \approx 1 \times 10^{-5}$), meaning that if they start coplanar, they remain coplanar and precess together.

Transit observations is also possible to put constraints on the minimum mutual inclination. If HD~93963~Ab and Ac showed a large lower limit, e.g., $\Delta I_{\rm{b,c;min}} \gtrsim 10^\circ$, that would suggest independent precession and a moderate true stellar obliquity ($\gtrsim 10^\circ$), while the observed value of $\Delta I_{\rm{b,c;min}} \sim 0.1^\circ$ \citep{Serrano2022} is too small to put any constraint.

\subsection{True obliquity and transit probability}\label{sec:obl-dyn}
\begin{figure*}
    \centering
    \includegraphics[width=0.7\linewidth]{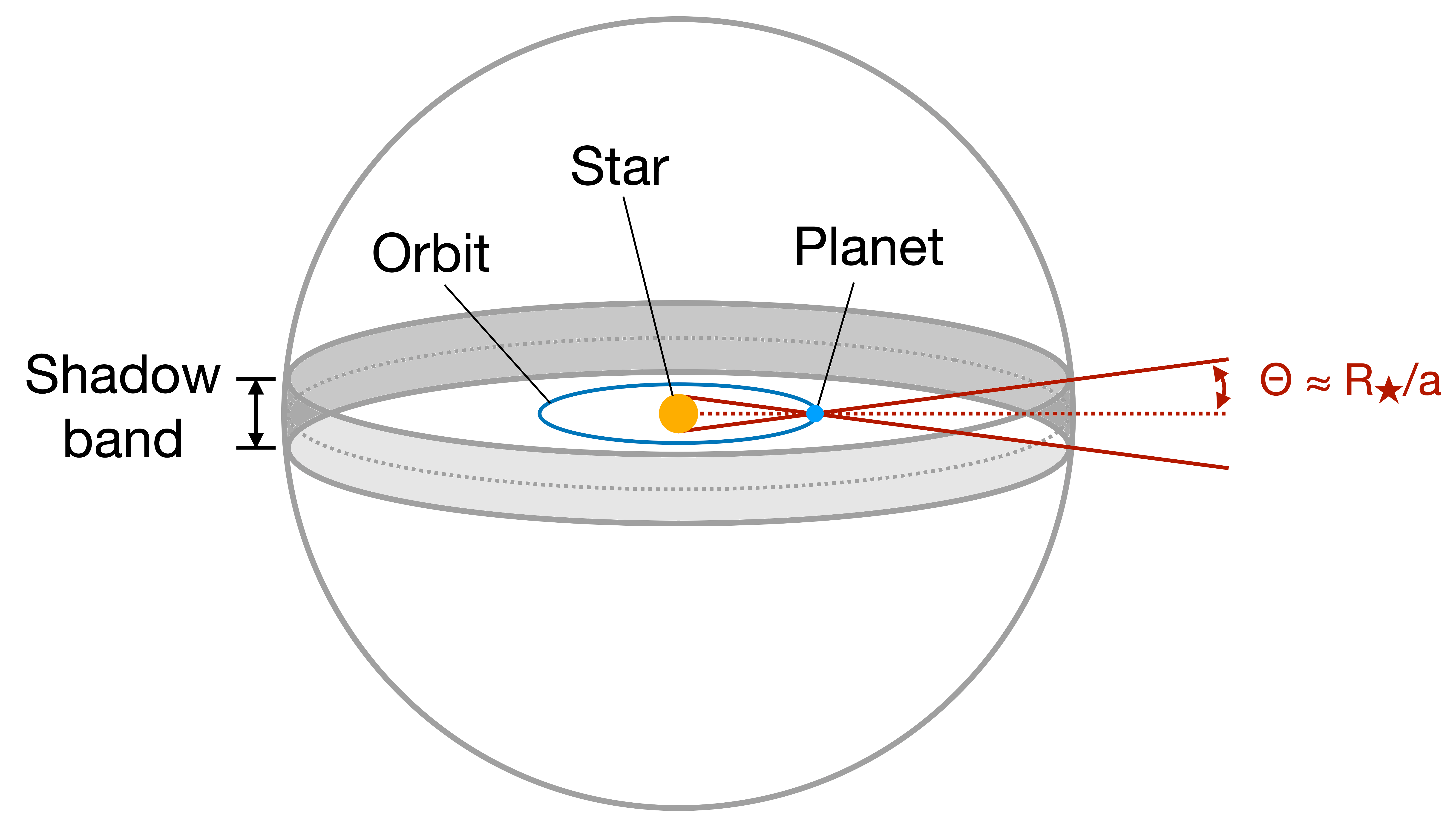}
    \includegraphics[width=0.48\linewidth]{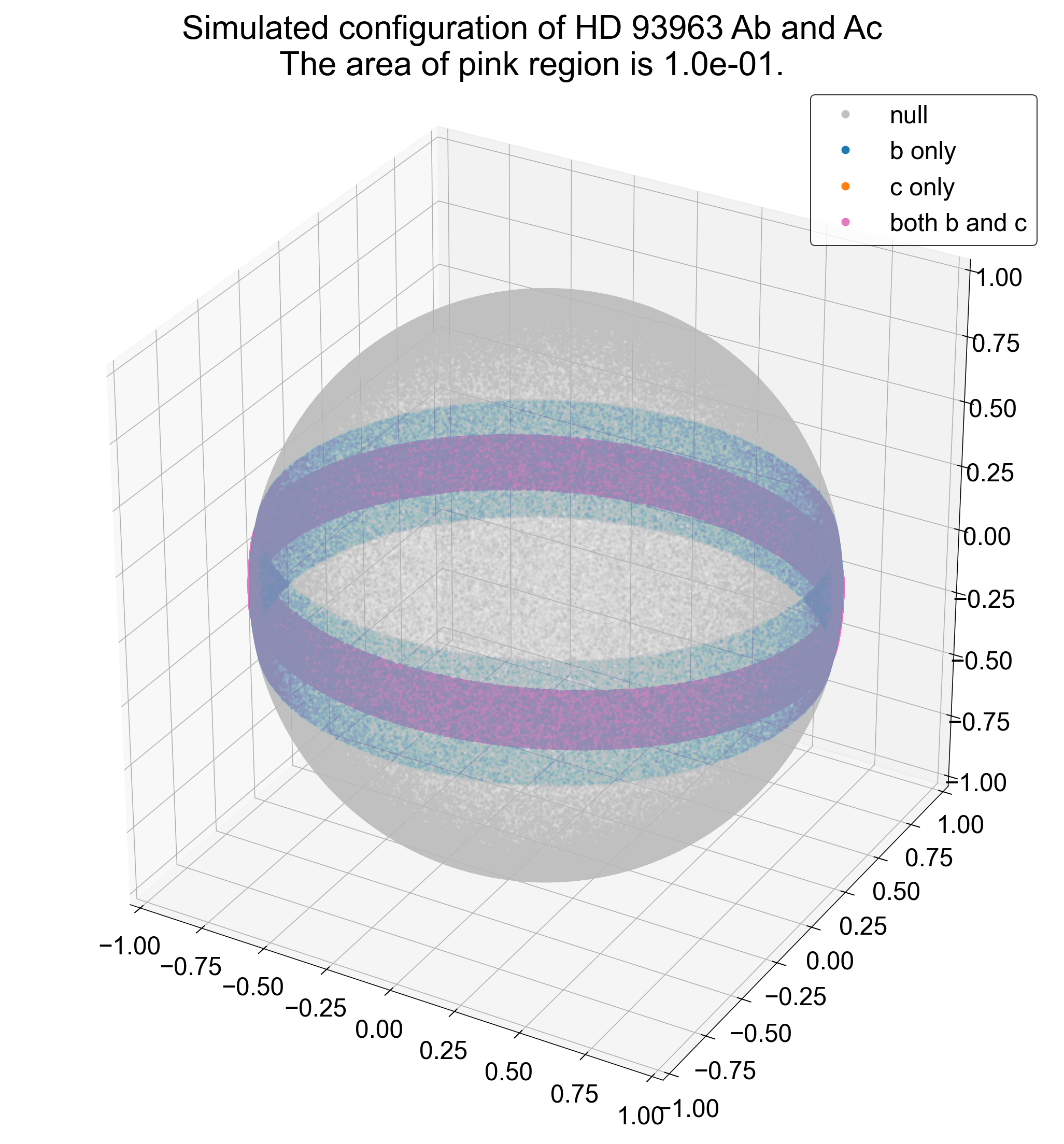}
    \includegraphics[width=0.48\linewidth]{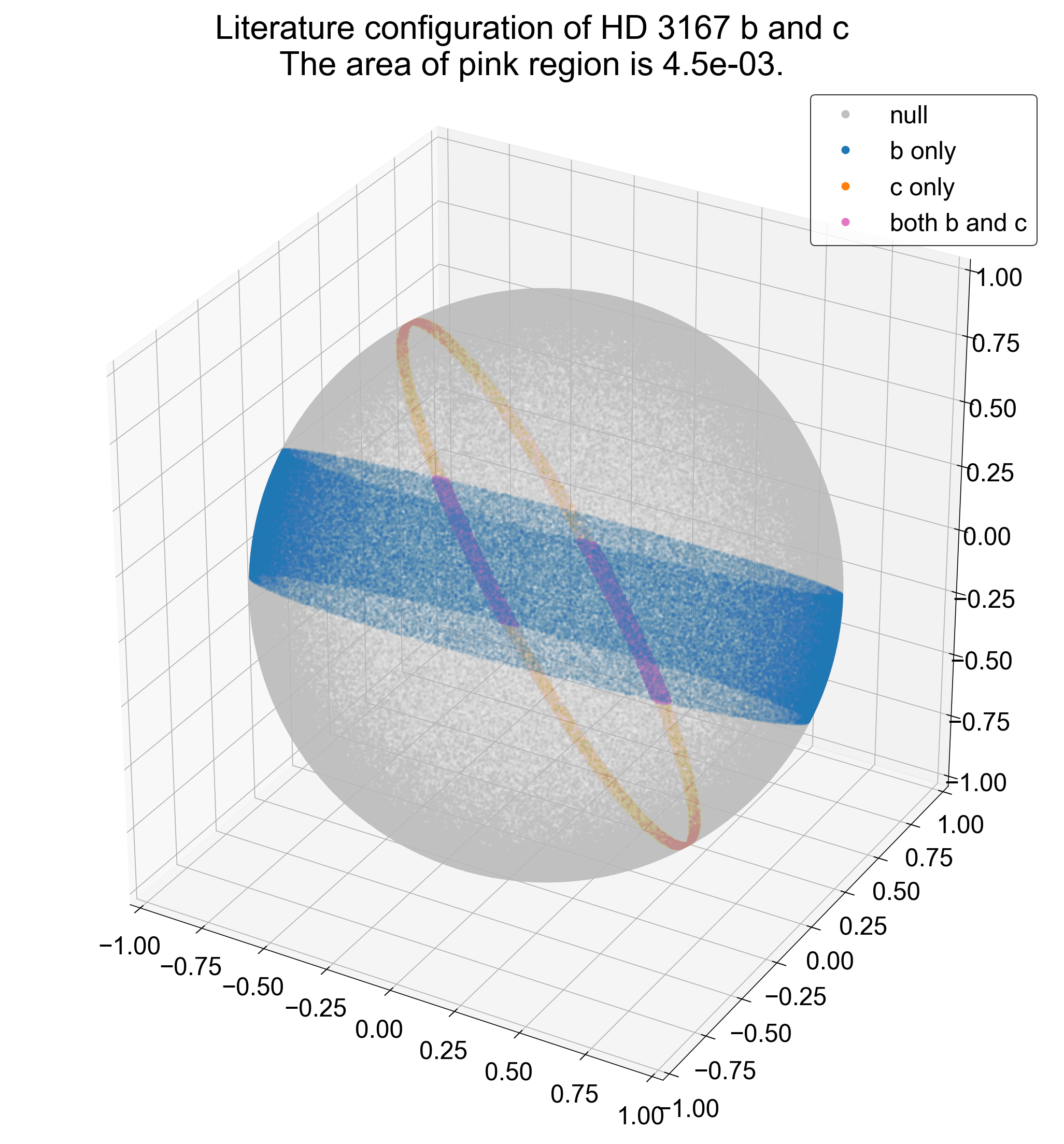}
    \caption{3D illustrations of transit shadow bands on the celestial sphere. Upper: A single transit toy model. The figure is modified from Figure 3 in \citet{Winn2010}, illustrating the geometry of a single transiting planet case. Lower: Cases that both planet can transit, with left one illustrating a simulated HD~93963~Ab and Ac, and right one illustrating HD~3167 b and c. The blue band and orange band respectively show two transit shadow bands of two transiting planets. The pink region shows the overlapped region, indicating the region where transit of both planets can be observed. The area of pink region of HD~3167 literature configuration is 0.0045.}
    \label{fig:prob}
\end{figure*}

Below, to explore how transit observation affected by true stellar obliquity, we quantify two possibilities: (1) the time-averaged probability that both HD~93963~Ab and Ac transit can be detected across the entire celestial sphere as a function of the initial true stellar obliquity, and (2) the time-varying probability for an observer at a fixed position to detect both planets transiting their host star.

We make two initial assumptions: (1) the planets are in circular orbits, and (2) their radii are negligible. Under these assumptions, we define a unified celestial sphere of radius $r = 1$, with the host star at its center. Each planet creates a shadow band on this sphere, within which its transit can be observed. The half-width of each band is $R_{\star}/a$ where $R_{\star}$ is the stellar radius, and $a$ is the planet's orbital semi-major axis. Figure~\ref{fig:prob} illustrates this geometry, and \citet{Winn2010} provides a detailed derivation. Consequently, for a single planet, the transit probability across the entire celestial sphere is simply $p = R_{\star}/a$. When multiple planets are present, the total transit probability is not simply the product of their individual probabilities \citep[$p \neq \prod_{i} p_i$, derivation can be found in e.g.,][\texttt{CORBITs} code]{Ragozzine2010, Brakensiek2016}. With at least three planets, the bands might not share a common overlap region. 

Hence, we use a Monte Carlo approach for simplicity. First, we uniformly sample $N$ points on the celestial sphere. We determine how many of these points lie within the shadow band of the first planet and then repeat the procedure for the second planet, restricting the sample to points already in the first planet’s band. For systems with more planets, we continue until all have been tested.

Finally, we count the number of points $N_{\rm{overlap}}$  that lie within all shadow bands. The probability of detecting multiple transits across the entire celestial sphere at a specific time is then $p = N_{\rm{overlap}} / N$. Notably, if $N_{\rm{overlap}} = 0$ in systems with three or more planets, it indicates no observer can detect every transit from any single location on the celestial sphere at that time.

\begin{figure*}[ht]
    \centering
    \includegraphics[width=0.49\linewidth]{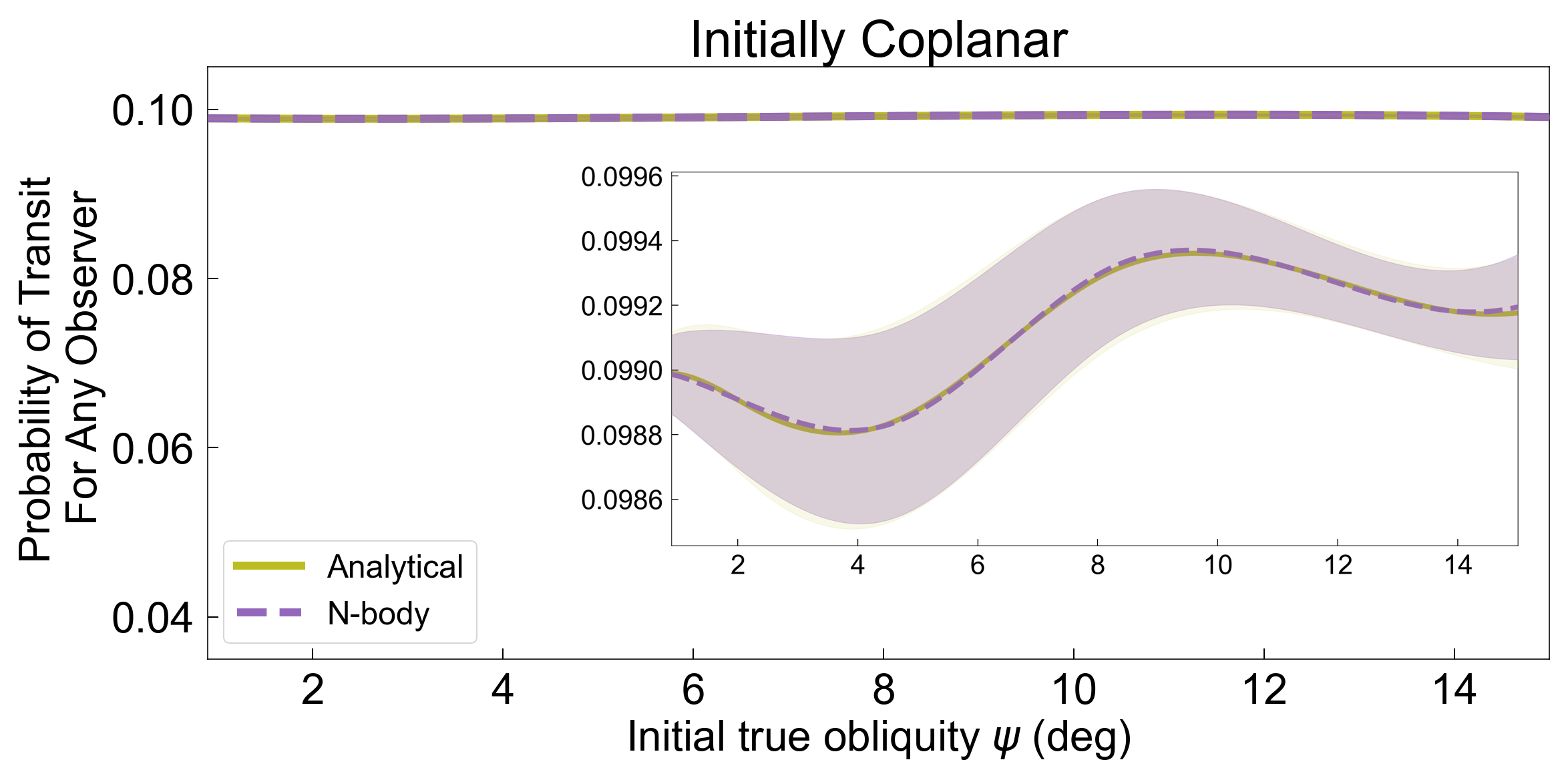}
    \includegraphics[width=0.49\linewidth]{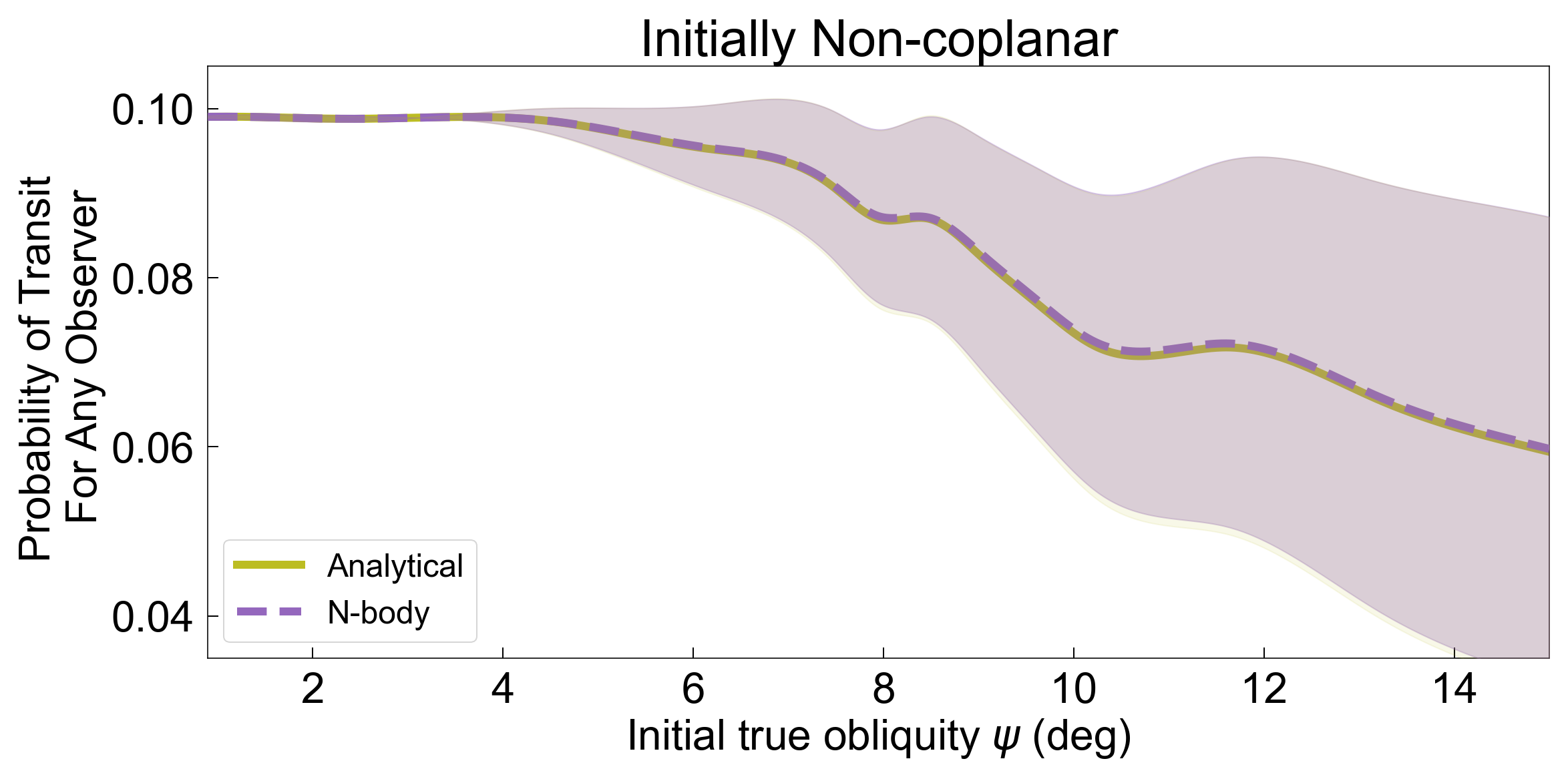}
    \caption{The the estimated probability of both HD~93963~Ab and Ac transiting their host star on the whole celestial sphere as a function of the initial true stellar obliquity ($\psi$). The left and right panels are respectively initialized from coplanar and non-coplanar configuration. The purple and olive colors represent results from the N-body simulation and the analytical solution, respectively.}
    \label{fig:mut_t_prob_sphere}
\end{figure*}
\begin{figure*}[ht]
    \centering
    \includegraphics[width=0.49\linewidth]{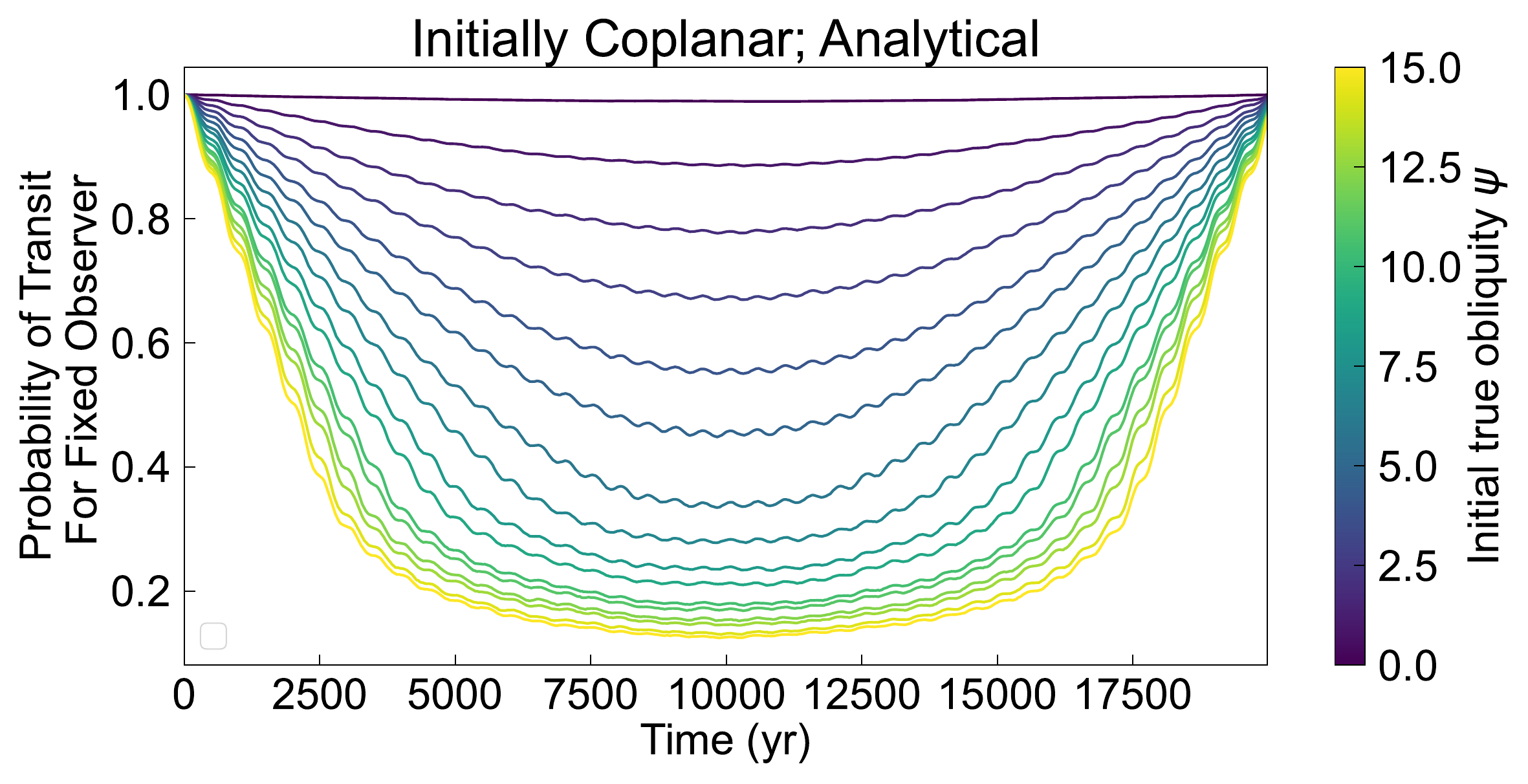}
    \includegraphics[width=0.49\linewidth]{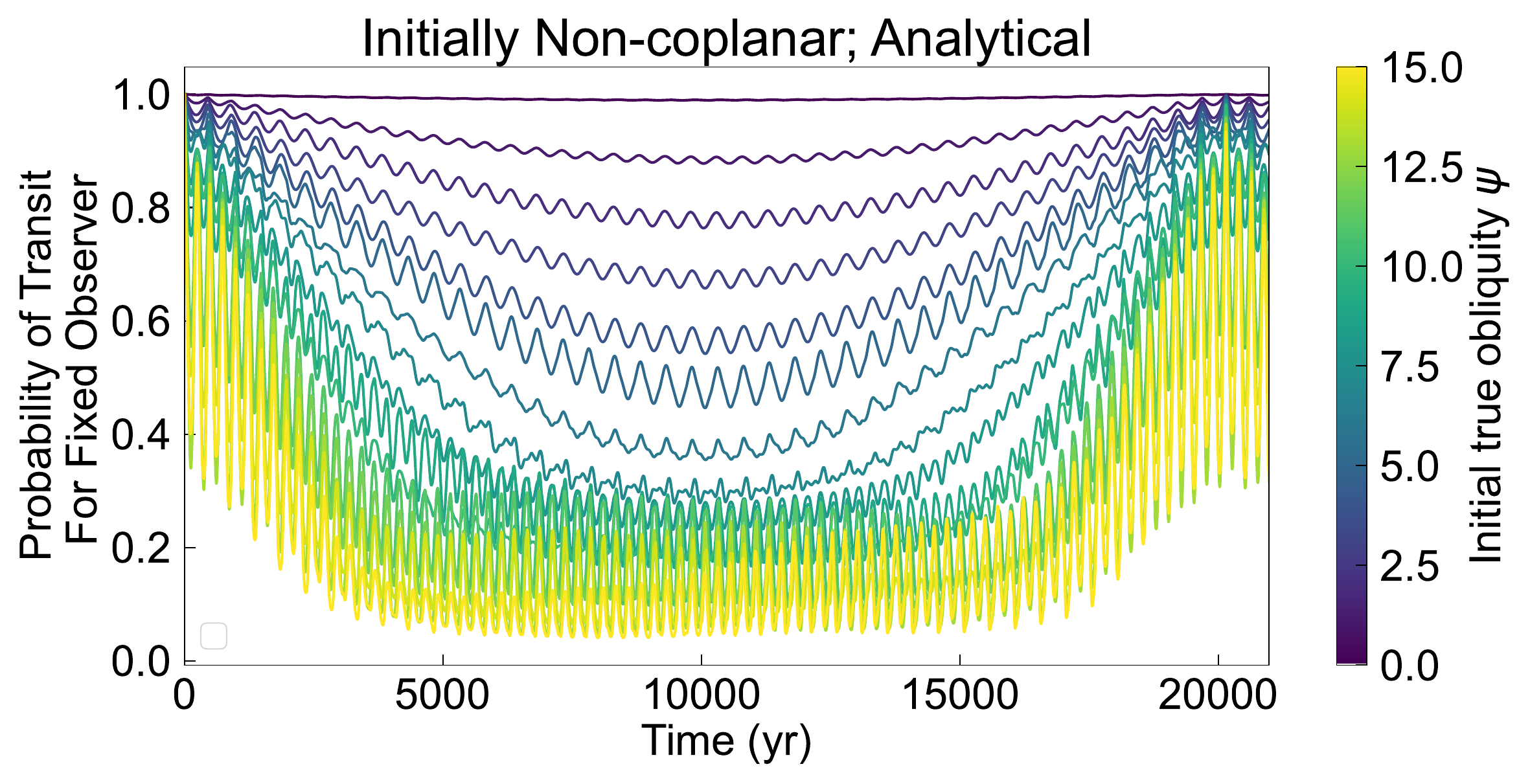}
    \caption{Time-varying probability that an observer at a fixed position could detect both HD~93963~Ab and Ac transiting their host star. The left and right panels are coplanar and non-coplanar initialization respectively obtained from analytical solutions. The color of the solid lines indicate the initial true stellar obliquity. A comparison derived from N-body simulations is shown in Figure \ref{fig:mut_t_prob_fix_append} in Appendix. }
    \label{fig:mut_t_prob_fix}
\end{figure*}

Based on the above methodology, we computed two main  probabilities. First, we determined the time-averaged probability that a randomly located observer could see both planets transit, as a function of the initial true stellar obliquity. For each simulation run, including the analytical solution and N-body simulations (as comparison), we calculated a time-averaged probability. Repeating this for various initial true stellar obliquities yielded the time-averaged probability as a function of the initial obliquity.

We tested both under coplanar and non-coplanar initializations (Figure \ref{fig:mut_t_prob_sphere}). The results from both analytical solution and N-body simulations match closely, consistent with our earlier findings on mutual inclination (Figure \ref{fig:mut_inc_inc}). For coplanar initialization, the time-averaged probability remains largely unaffected by the initial true stellar obliquity. However, for non-coplanar initialization, it declines noticeably above a threshold of roughly $4^{\circ}$ for HD~93963 system. Thus, the observed transits of planets b and c suggest the two planets are likely coplanar or, if not, that they have low obliquity while precessing independently.

Next, we calculated the time-varying probability that a fixed-position observer could see both HD~93963~Ab and Ac transit their host star as a function of the initial true stellar obliquity. We determined the overlap between the transit regions on the celestial sphere at $t=0$ and at later times $t = t_{i}$. By repeating this procedure over multiple time steps and averaging across multiple runs for each obliquity, we obtained the time-varying probability at a fixed observer’s location.

Figure \ref{fig:mut_t_prob_fix} shows the time-dependent probability for coplanar and non-coplanar configurations. Both exhibit a duty cycle of about 20 kyr, during which the transit probability drops below $\sim$20\% if the initial obliquity exceeds 10 degrees. For non-coplanar initialization above 10 degrees, the probability follows a similar 20-kyr pattern as coplanar initialization, but also declines more rapidly to $\lesssim$50\% near the start and end of this cycle, following a shorter $\sim$245-year cycle (about half the main eigenperiod). This behavior makes higher obliquities in non-coplanar configurations less likely.

Thus, considering both the celestial-sphere and fixed-observer perspectives, the true stellar obliquity of HD~93963~Ab and Ac should be only a few degrees, lower than but within the observed sky-projected stellar obliquity of $\lambda = 14^{+17}_{-19}$ degrees.

In summary, we present a methodology for dynamically estimating the true stellar obliquity and apply it to the inner two planets of HD~93963 system. Our analysis indicates that the orbits of HD~93963~Ab and Ac are most likely coplanar with a small mutual inclination, and are well aligned with the stellar spin. This approach can be readily extended to other systems satisfying the Laplace-Lagrange model.

\section{Discussion}\label{sec:discuss} 
\begin{figure*}[ht]
    \centering
    \includegraphics[width=0.9\linewidth]{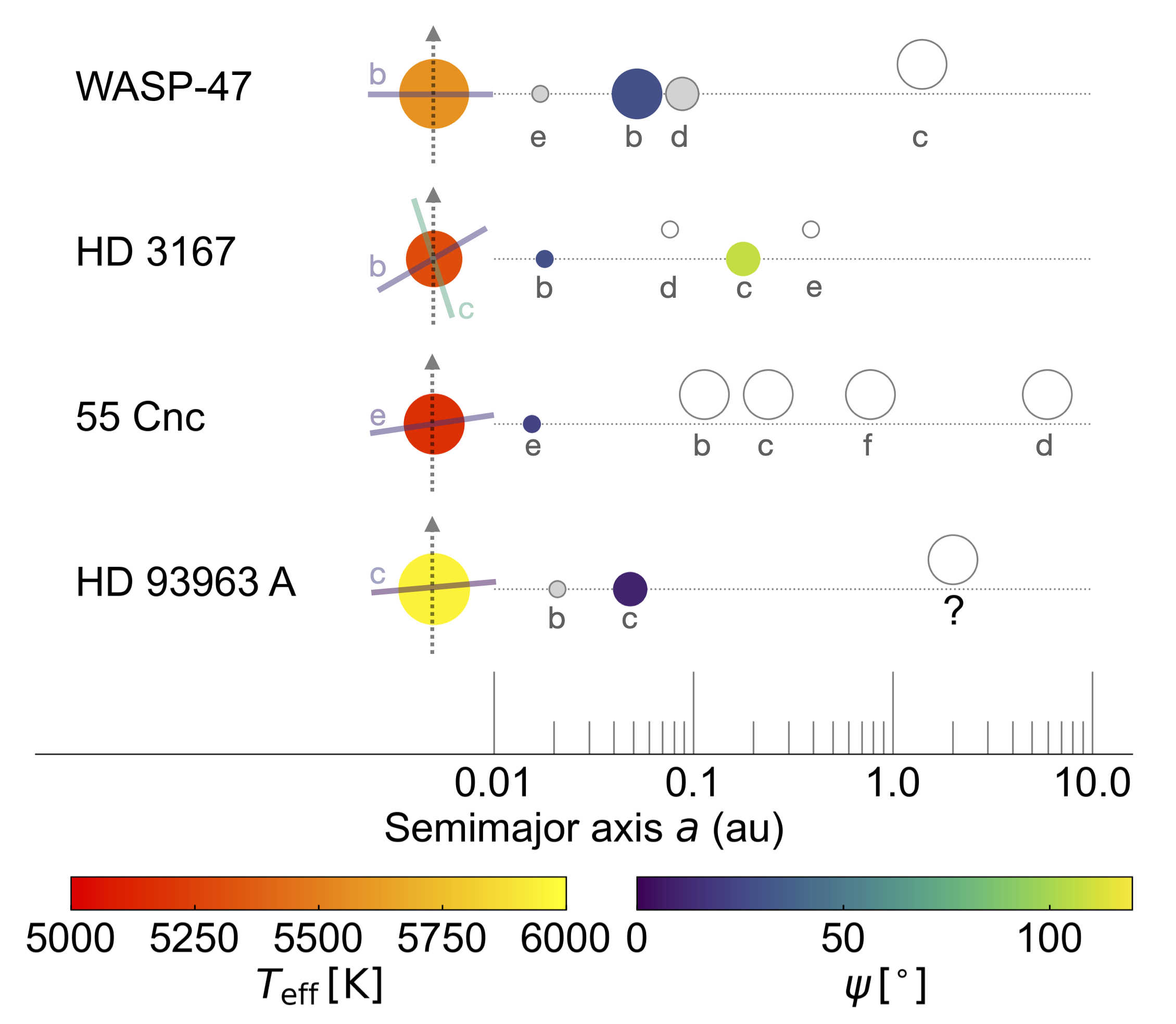}
    \caption{Planetary systems containing ultra-short-period planets (USPs) and having measured obliquities. Host stars are represented by filled circles, with their effective temperatures indicated by color. The stellar spin orientation and orbit planes are marked by dotted arrows and solid lines, respectively. Planets are depicted as circles on the right part of the figure, where larger circles correspond to more massive planets. The color of each circle reflects the true stellar obliquity from observations, while gray circles represent transiting planets without obliquity measurements, and white circles represent non-transiting planets. Notably, the outer cold giant planet of HD~93963 system has not been confirmed.}
    \label{fig:usp}
\end{figure*}
\begin{figure*}[ht]
    \centering
    \includegraphics[width=0.46\linewidth]{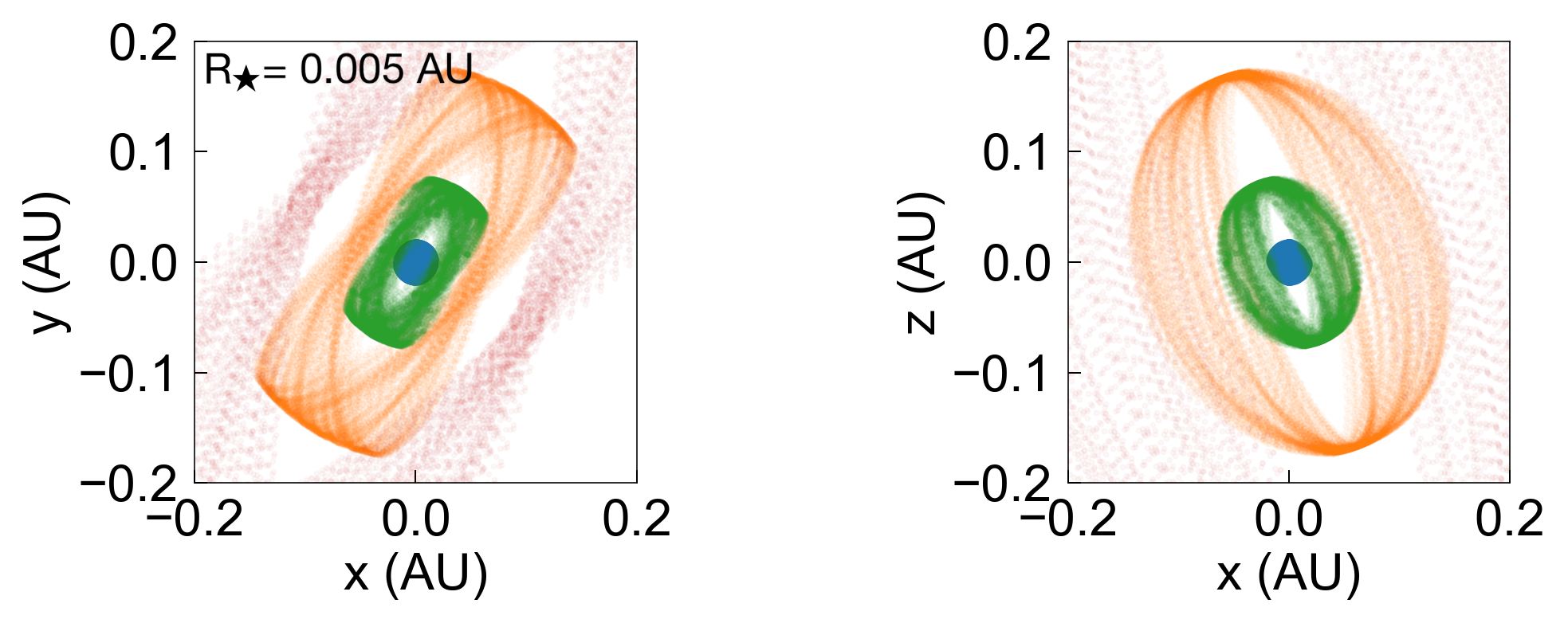}
    \includegraphics[width=0.49\linewidth]{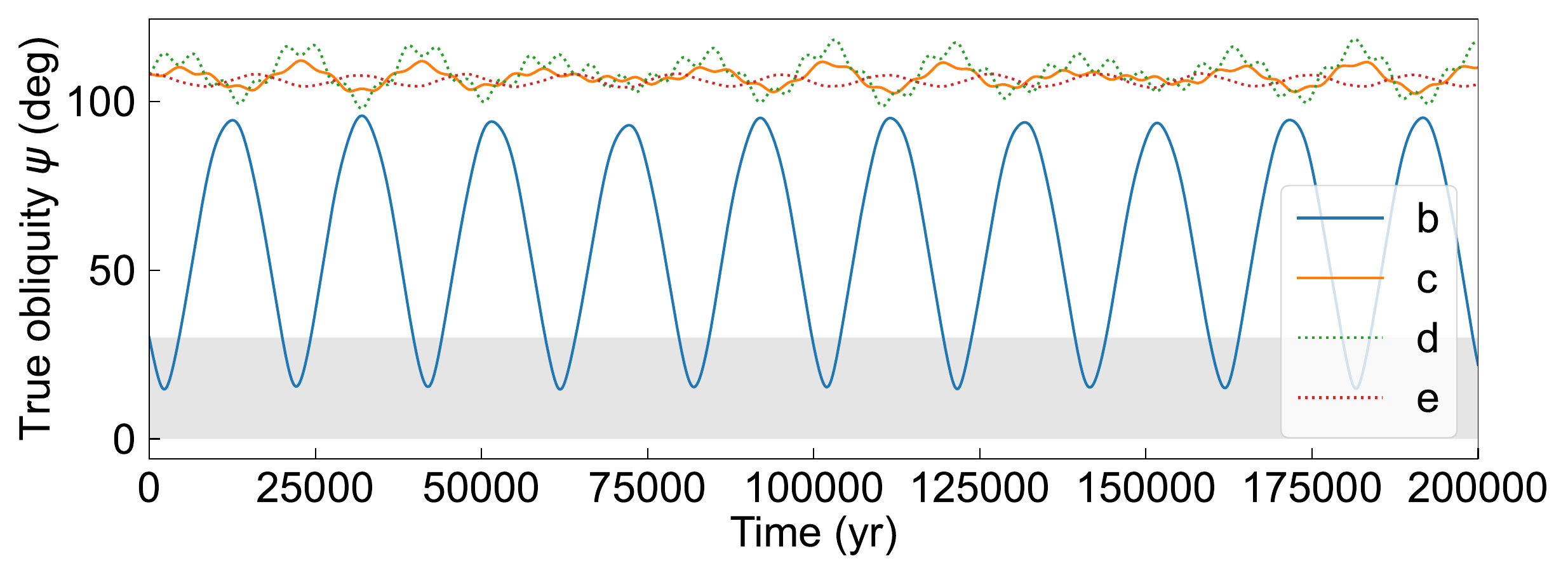}
    \includegraphics[width=0.46\linewidth]{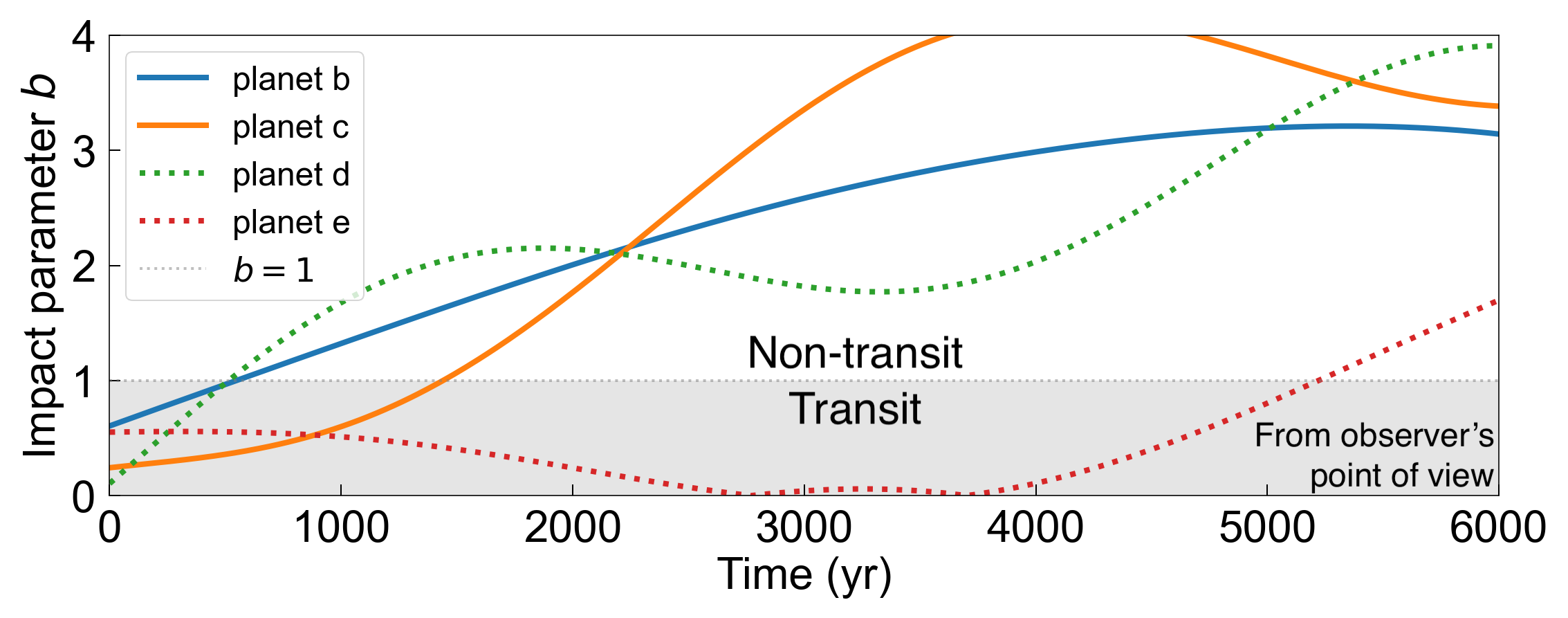}
    \includegraphics[width=0.49\linewidth]{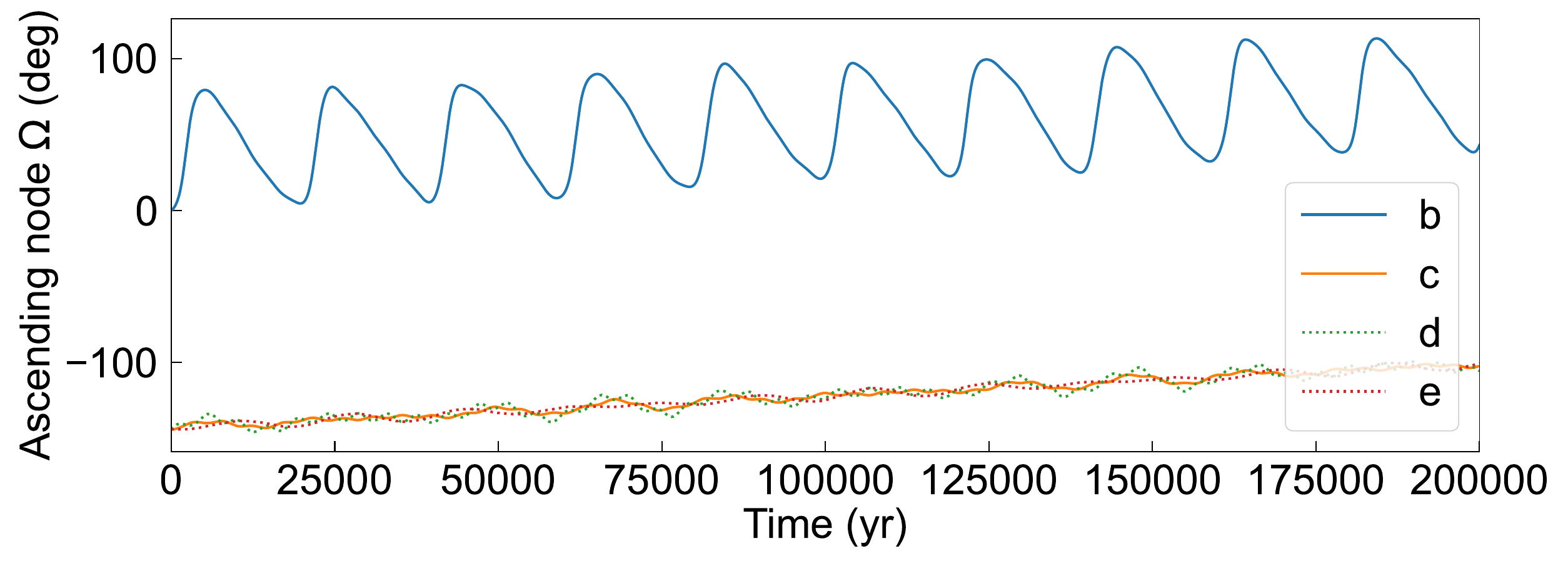}
    \caption{A randomly selected 200-kyr stability test of the HD~3167 system having a configuration that planets d, c, and e are initially coplanar, while planet b is significantly inclined relative to the other three planets. Upper left: the position of all planets during the 200-kyr simulation in Cartesian coordinate. Upper right: The evolution of true stellar obliquity ($\psi$). The shaded region illustrates true stellar obliquity of planet b lower than 30 degrees. Lower right: The evolution of ascending node ($\Omega$). Lower left: The 6-kyr evolution of the impact factor for all planets for an observer positioned at a random and fixed location where transits of HD~3167 b and c are detectable at time equal to zero. The shaded regions indicate the time during which the selected observer can detect the planets transiting the host star. }
    \label{fig:HD3167_sim}
\end{figure*}
\begin{figure*}[ht]
    \centering
    \includegraphics[width=0.80\linewidth]{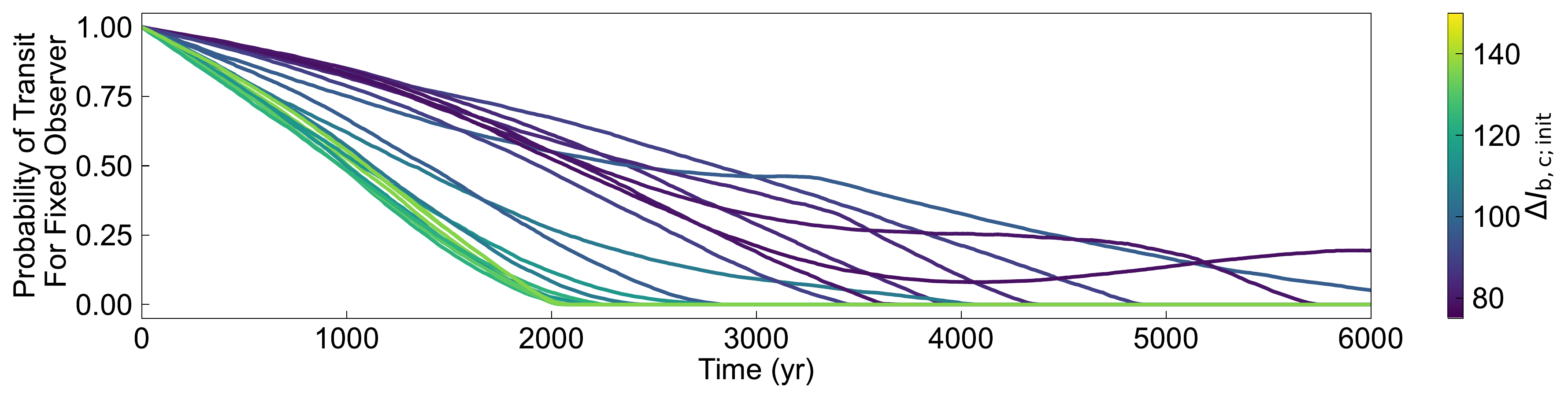}
    \caption{Time-varying probability that an observer at a fixed position could detect both HD~3167 b and c transiting their host star. The system is initialized from the configuration that planets d, c, and e are initially coplanar, while planet b is significantly inclined relative to the other three planets. The first 6 kyr is shown and different colors in both panels represent results from various initial conditions.}
    \label{fig:HD3167_prob}
\end{figure*}

\subsection{USP systems and stellar obliquity}
The HD~93963 system contains an ultra-short-period (USP) planet with $R_{\rm{p}}=1.35\,R_{\oplus}$ in a $P=1.04\,\mathrm{d}$ orbit around a relatively bright host star ($V=9.2$). It is also the fourth USP-hosting system for which a sky-projected stellar obliquity $\lambda$ measurement is available. However, this measurement is not a direct determination of the USP’s true stellar obliquity but is inferred through its coplanar planet and the system’s overall dynamics. Directly measuring a USP’s stellar obliquity remains technically challenging, chiefly because the RM effect amplitude is typically below $1\,\mathrm{m\,s}^{-1}$. The amplitude can be estimated using the following empirical equation:
\begin{equation}
    A_{\rm{RM}} = \sqrt{1-b^2} \times \left( \frac{ R_{\rm{p}} }{R_{\rm{s}}} \right)^2 \times v \sin i.
\end{equation}
For instance, HD~93963~Ab has an impact parameter of $b=0.2$, a planet-to-star radius ratio of $R_{\mathrm{p}}/R_{\mathrm{s}}=0.012$, and a projected stellar rotational velocity of $v \sin i = 2.96\,\mathrm{km\,s}^{-1}$. This combination yields an RM signal amplitude of only $A_{\mathrm{RM}} = 0.4\,\mathrm{m\,s}^{-1}$. Such a low amplitude can be overwhelmed by stellar noise sources, such as oscillations, granulation, and starspots, even with the best RV instruments (e.g., Keck/KPF, Gemini/Maroon-X, VLT/ESPRESSO), which achieve spectral resolutions over 80,000 and precisions of $\sim0.3\,\mathrm{m\,s}^{-1}$. It remains uncertain whether techniques for mitigating stellar noise, such as Gaussian Process (GP) modeling, can reliably recover the tiny RM signal. Moreover, a USP typically has a transit duration of under 1.5~hours, restricting the achievable time resolution. Consequently, only bright USP-hosting stars, where exposure times of a few minutes or less are feasible, serve as promising targets for measuring the RM effect of USPs.

So far, only four USP-hosting systems have sky-projected stellar obliquity $\lambda$ measurements on any of the planets within the systems: WASP-47, HD~3167, 55~Cnc, and HD~93963~A. A common feature among these is that the USPs appear spin-orbit aligned, and the host stars are older than 1~Gyr. Beyond that, their planetary system configurations differ substantially (Figure~\ref{fig:usp}). Notably, HD~3167 hosts a warm mini-Neptune on a highly misaligned orbit. This diversity suggests that each system may have followed a distinct formation pathway. Below, we briefly discuss WASP-47 and 55~Cnc, while focusing on HD~3167 for its unusual orbital configuration. 

\subsection{WASP-47 and 55 Cnc}
\textit{WASP-47} --- This system contains a USP orbiting interior to a well-aligned hot Jupiter (planet b; \citealt{Sanchis-Ojeda2015, Bourrier2023}), along with a smaller planet (planet d) and another giant planet (planet c). An \textit{in situ} formation scenario might explain the origins of both the USP and the hot Jupiter in this system \citep{Huang2016}. Or alternatively, the USP could have formed \textit{in situ} in a gas-poor region after the hot Jupiter migrated to its current position following formation in a gas-rich disk \citep{Weiss2017}.

\textit{55 Cnc} --- This system hosts a USP with $\psi=23_{-12}^{+14\circ}$, measured directly via the RM effect \citep{Zhao2023}, along with four non-transiting giant planets located beyond 0.11 AU. \citet{Zhao2023} state that the spin-orbit alignment of 55 Cnc e supports a dynamically gentle migration pathway, likely involving tidal dissipation through low-eccentricity planet-planet interactions and/or planetary obliquity tides. Additionally, the outermost planet has a minimum mass of about $3\,M_{\rm{J}}$ and an orbital period of roughly 5000 days \citep{Rosenthal2021}. A joint analysis of RV and astrometry could further characterize its orbital inclination, offering additional insight into the formation history of the 55 Cnc system. 

\subsection{HD~3167}
This system includes an USP with a reported true stellar obliquity $\psi_{\rm{b}}=29.5_{-9.4}^{+7.2\circ}$, directly measured using the RM effect revolutions technique \citep{Bourrier2021}, as well as three additional warm Neptune-like planets within 1 AU. The third inner planet (planet c) also transits, and was reported to have a polar orbit with $\psi_{\rm{d}}=107.7_{-4.9}^{+5.1\circ}$ \citep{Dalal2019,Bourrier2021}. 

We tested the reported the time evolution of system architecture using the N-body simulation framework described in Section \ref{sec:nbody-comp}. Notably, the Laplace-Lagrange theory (Section \ref{sec:ana_sol}) is not applicable for such a large true mutual inclination ($\sim$102$^{\circ}$) between planets b and c.

We performed 20 independent runs with an integration time of 200 kyr, which is sufficient to capture the relevant dynamics. For simplicity, we assumed planets d and e are initially coplanar with transiting planet c. Because planet d and e have relatively long orbital periods, even a small true mutual inclination could prevent them from transiting. We then assumed planet b is significantly inclined relative to the other three planets following \citet{Bourrier2021}.  We agnostically set the differences of ascending nodes between planet b and outer planets to be uniformly distributed from 0 to 360 degrees. All 20 runs remained stable over 200 kyr, and a representative example is shown in Figure \ref{fig:HD3167_sim}.

Overall, due to secular perturbations from the outer planets, planet b cannot maintain its transiting orientation or the reported true stellar obliquity. We found that the fraction of time when planet b has $\psi < 30^{\circ}$ is only about 10\%. The probability of planets b and c both transiting the entire celestial sphere ranges from $\sim$22\% down to $\sim$0.3\%. In other words, the {\it K2} mission was ``quite lucky'' to observe the system in a high-inclination yet both-transiting configuration, and thus the existing results were questioned.

Moreover, as shown in the lower left panel of Figure \ref{fig:HD3167_sim}, the impact parameters of planets b and c still evolve quickly with large amplitudes in this configuration. On century-long timescales, one planet is likely to become non-transiting. On shorter, decade-long timescales, the changing impact factors should produce substantial transit duration variations for HD~3167 b and c, which future missions such as ESO/PLATO \citep{Rauer2014} could test. We also encourage additional Rossiter-McLaughlin observations to confirm the results reported by \citet{Bourrier2021}.

\section{Conclusions}\label{sec:conclusion}
In summary, we observed the Rossiter-McLaughlin effect of a mini-Neptunian planet, HD~93963~Ac, accompanied by an inner ultra-short-period super-Earth—on May 3rd, 2024 UT with Keck/KPF. Our analysis revealed a spin-orbit alignment, yielding a sky-projected stellar obliquity of $\lambda_{\rm{c}}=14^{+17\circ}_{-19}$. 

Long-term radial velocity (RV) data suggests the presence of an outer giant planet companion in the HD~93963 system (\citealt{Serrano2022} and \citealt{VanZandt2025}). We performed dynamical analyses using Laplace-Lagrange secular theory and N-body simulations under a three-planet configuration. Our calculations of the true mutual inclination between HD~93963~Ab and Ac as a function of the true stellar obliquity, along with transit probability estimates for these inner planets, indicate that they are likely aligned with the stellar spin to within a few degrees.

HD~93963 system is the fourth system with an ultra-short-period planet (USP) and a measured stellar obliquity of any planet within the system, following WASP-47, HD~3167, and 55 Cnc, which exhibit diverse orbital architectures. While HD~93963, WASP-47, and 55 Cnc appear to favor largely coplanar orbits, HD~3167 has been reported to have a large true mutual inclination ($\sim$100$^{\circ}$) between planets b and c. However, our additional dynamical analysis suggests that differential nodal precession makes this reported configuration less likely and a low double-transit probability also challenges existing results.
We encourage further obliquity measurements of HD~3167 b and c to better constrain its orbital architecture.

\section*{Acknowledgements} 
This work is supported by the Korea Astronomy and Space Science Institute under the R\&D program (project No. 2025-1-830-05) supervised by the Ministry of Science and ICT, and National Key R\&D Program of China, No. 2024YFA1611802. 
H.Y.T. appreciates the support by the EACOA/EAO Fellowship Program under the umbrella of the East Asia Core Observatories Association. 
E.K.\ is supported by JSPS KAKENHI Grants No.\ 24K00698 and 24H00017. 
J.M.J.O. acknowledges support from NASA through the NASA Hubble Fellowship grant HST-HF2-51517.001-A, awarded by STScI. STScI is operated by the Association of Universities for Research in Astronomy, Incorporated, under NASA contract NAS5-26555. 
H.Y. acknowledges funding from the European Research Council under the European Union’s Horizon 2020 research and innovation programme (grant agreement No. 865624, GPRV).
This research has made use of the NASA Exoplanet Archive, which is operated by the California Institute of Technology, under contract with the National Aeronautics and Space Administration under the Exoplanet Exploration Program.
We thank the time assignment committees of the University of California, the California Institute of Technology, NASA, and the University of Hawai'i for supporting the {\it TESS}-Keck Survey with observing time at the W. M. Keck Observatory.
We gratefully acknowledge the efforts and dedication of the Keck Observatory staff for support of HIRES and remote observing.
We recognize and acknowledge the cultural role and reverence that the summit of Manua Kea has within the indigenous Hawaiian community. We are deeply grateful to have the opportunity to conduct observations from this mountain.
We thank Songhu Wang for helpful suggestions and comments to improve the paper.

\facility{Keck:I (KPF and HIRES), {\it TESS}, Exoplanet Archive}
\software{\texttt{numpy} \citep{numpy2020}, \texttt{scipy} \citep{SciPy-NMeth2020}, \texttt{batman} \citep{Kreidberg2015}, \texttt{emcee} \citep{Foreman-Mackey2013}, \texttt{EXOFAST} \citep{Eastman2013}, \texttt{isoclassify} \citep{Huber2017}, \texttt{gyro-interp} \citep{Bouma2023}, \texttt{SpecMatch} \citep{Petigura2015, Yee2017}.} 

\appendix
\section{Additional Figures}\label{sec:rm-appendix}
\begin{figure*}
    \centering
    \includegraphics[width=0.95\linewidth]{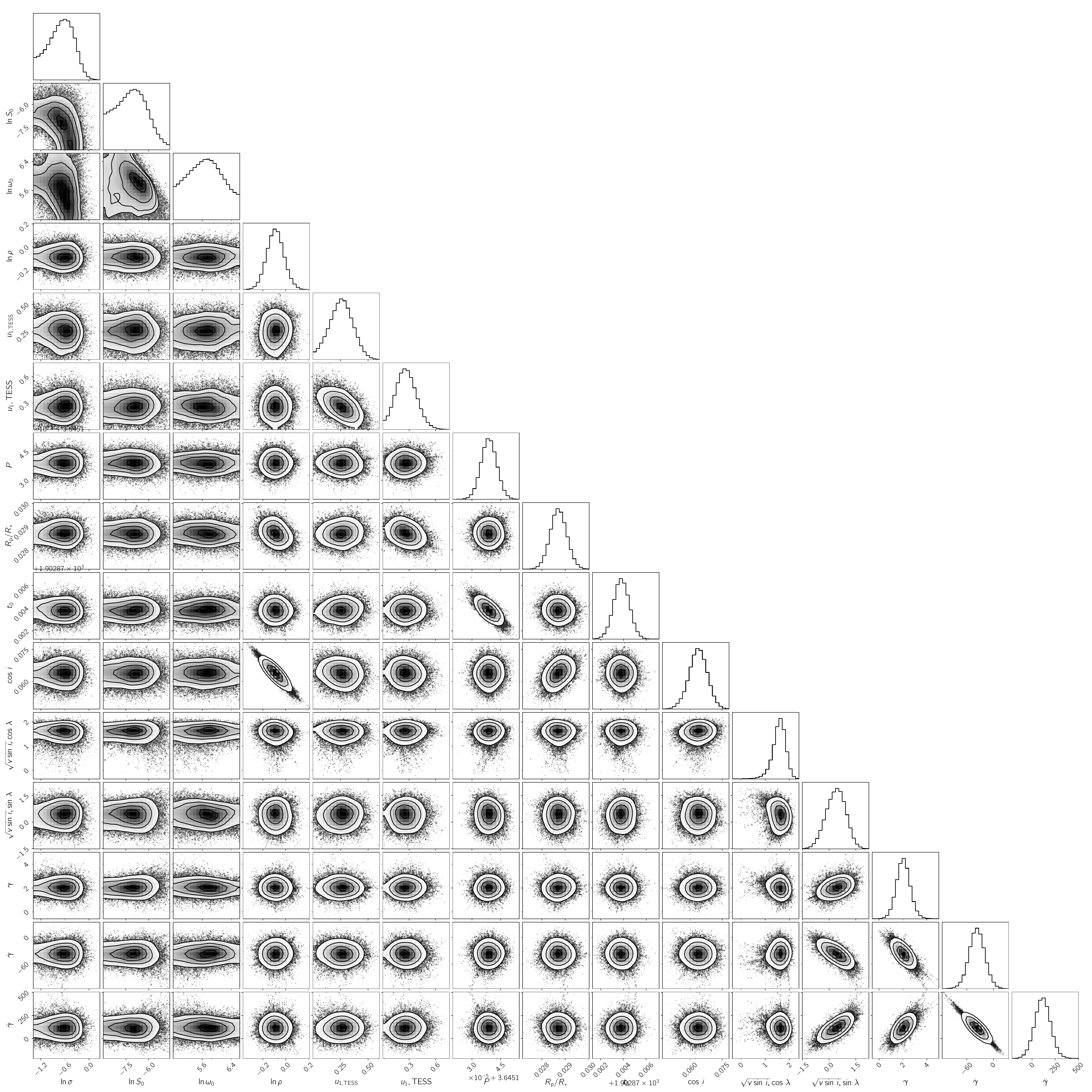}
    \caption{The posterior distribution of fitted parameters in Table \ref{tab:planet_para}, derived from MCMC sampling.}
    \label{fig:corner}
\end{figure*}

\begin{figure*}[ht]
\centering
    \includegraphics[width=0.495\linewidth]{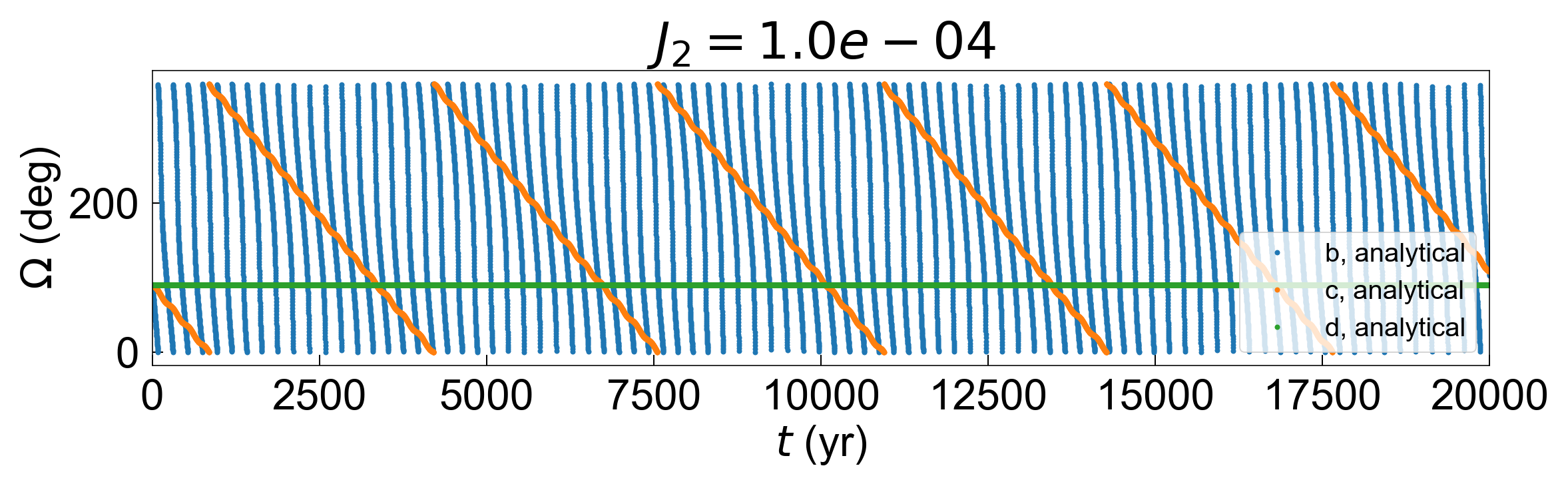}
    \includegraphics[width=0.495\linewidth]{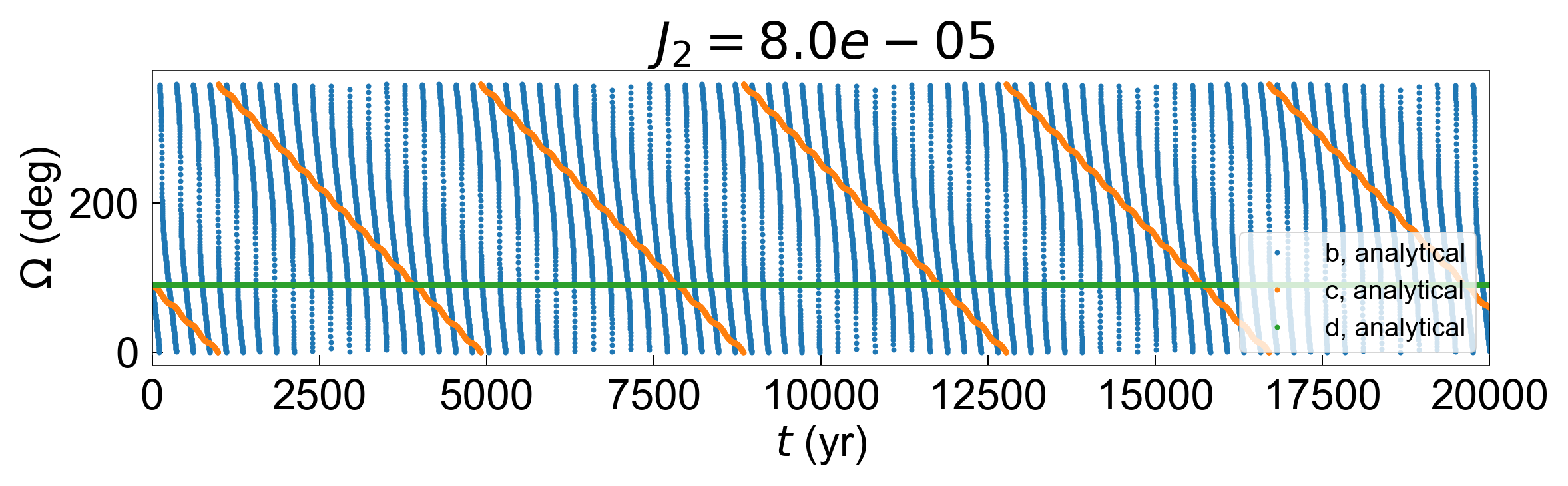}
    \includegraphics[width=0.495\linewidth]{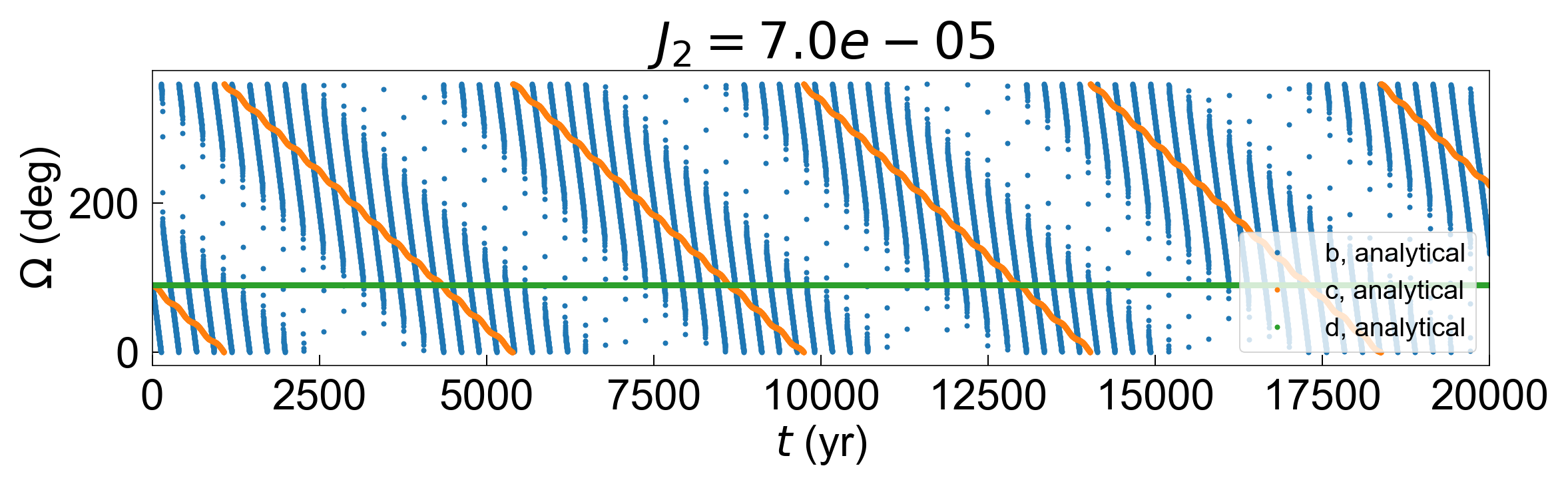}
    \includegraphics[width=0.495\linewidth]{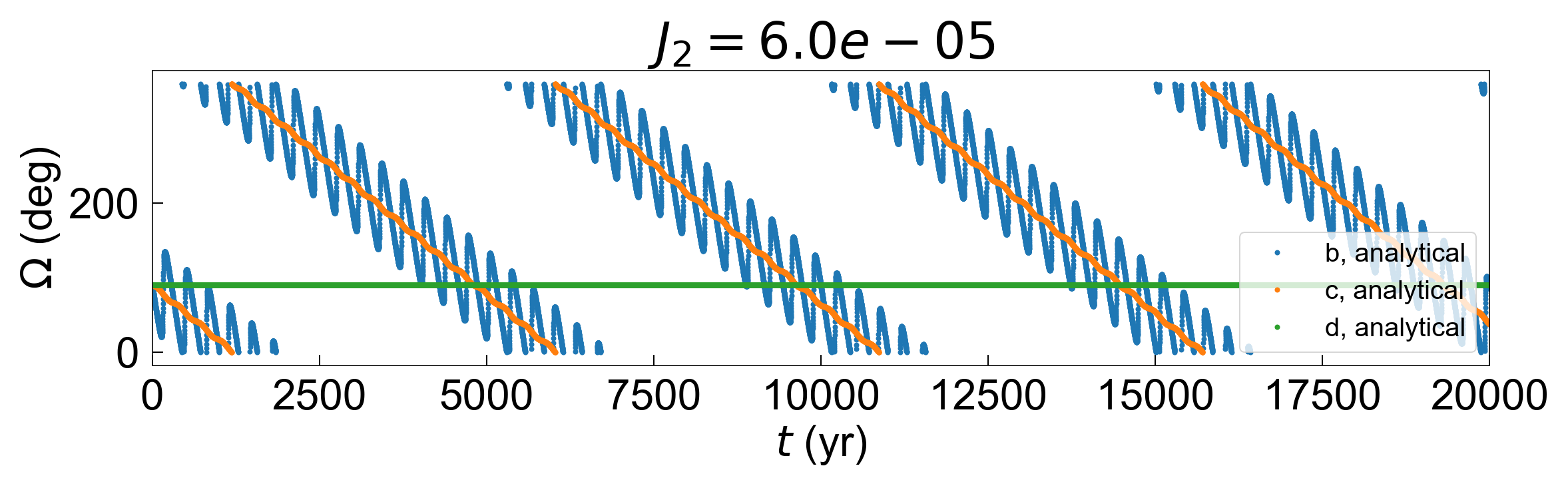}
    \includegraphics[width=0.495\linewidth]{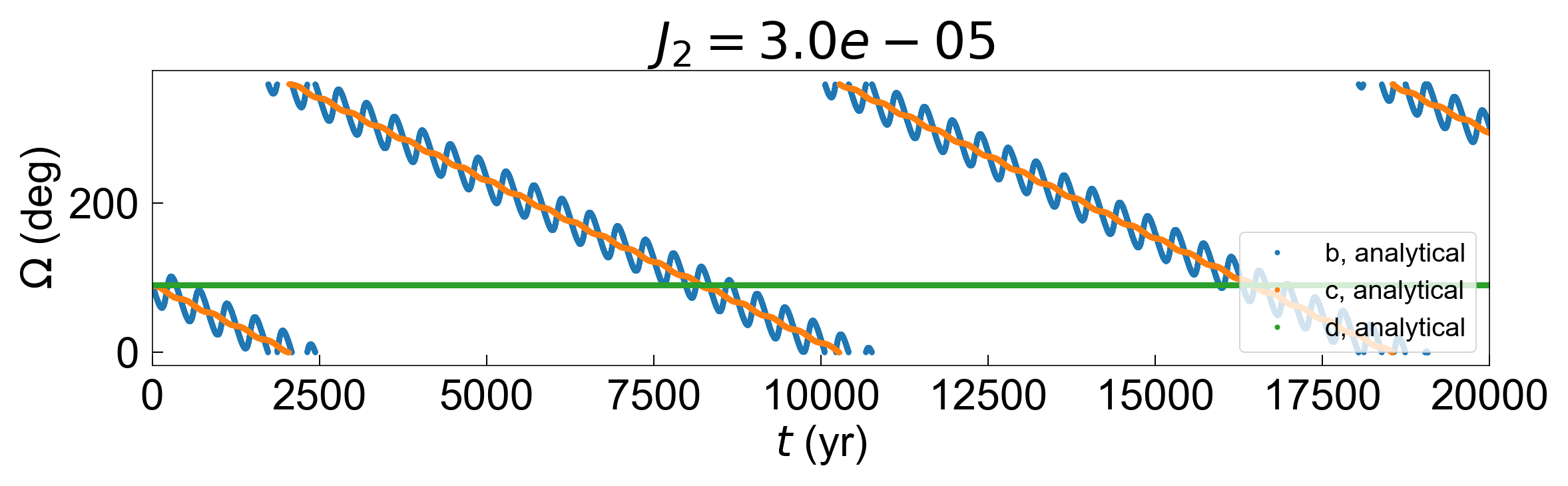}
    \includegraphics[width=0.495\linewidth]{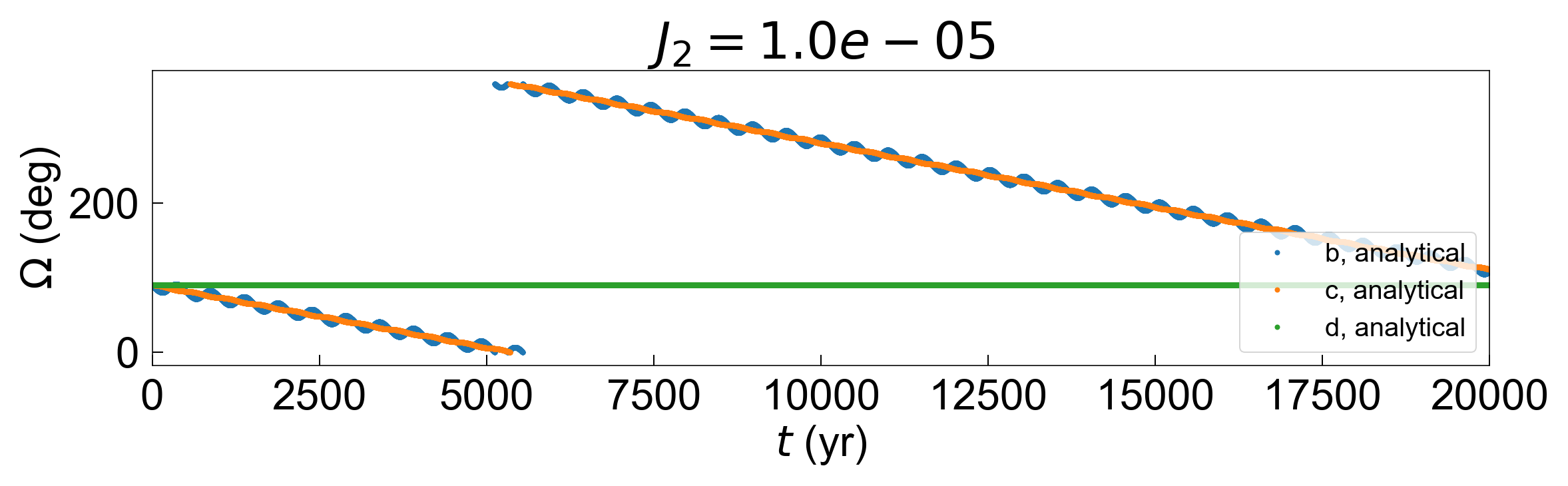}
    \caption{Nodal Precession varying by $J_2$ value for HD~93963 system as a supplementary to Figure \ref{fig:J2_variation}. The dominant precession rate of the planet b (USP) changes as $J_2$ value changes. Planet b and c precess separately at different eigenvalues by $J_2 \gtrsim 0.8\times10^{-5}$ while they precess together at a same eigenvalue as $J_2 \lesssim 0.6\times10^{-5}$. $J_2 = 1.0\times10^{-5}$, shown in the bottom right, represents the value we derived for the host star HD~93963~A.}
    \label{fig:J2_variation_appendix}
\end{figure*}

\begin{figure*}[ht]
    \centering
    \includegraphics[width=0.49\linewidth]{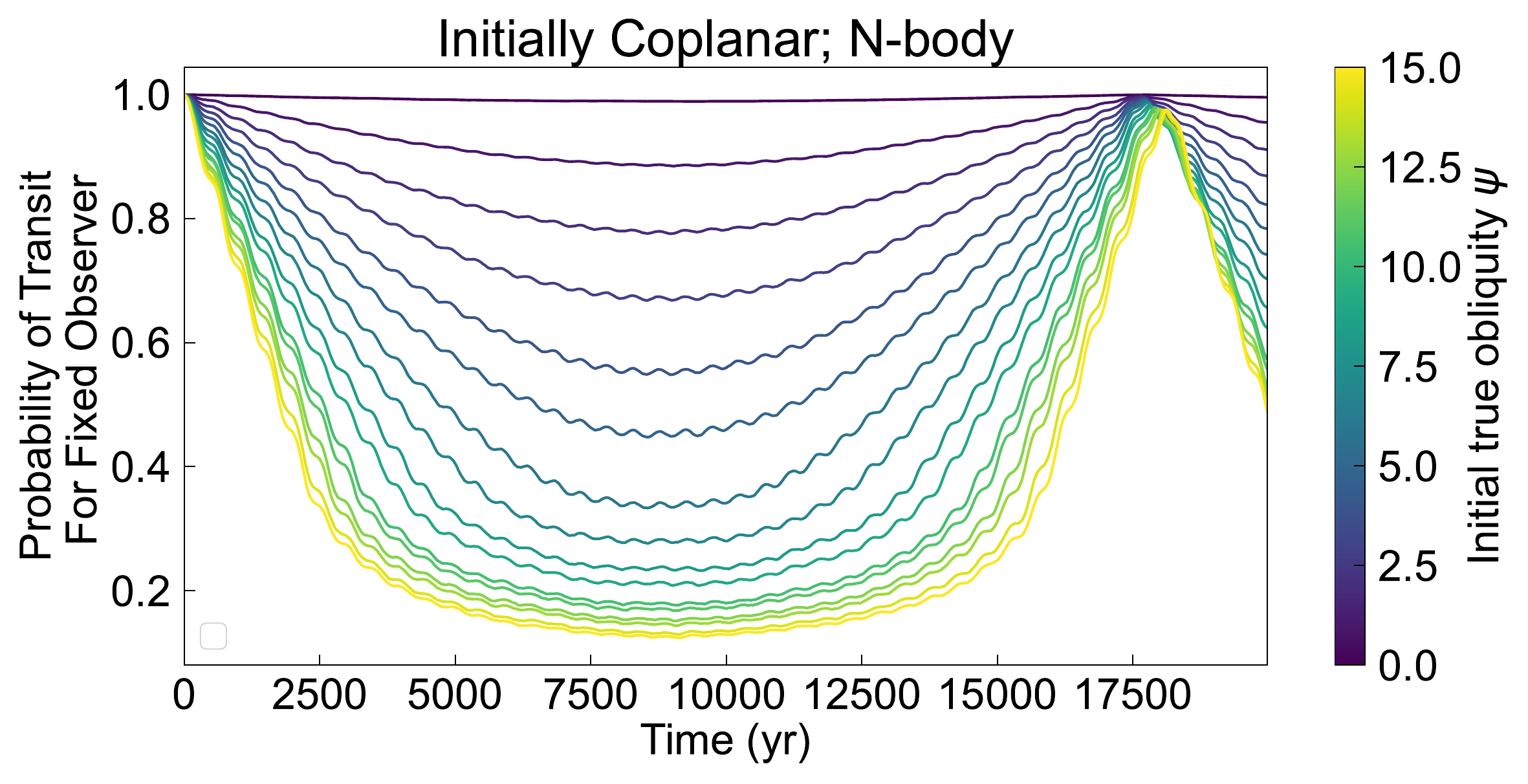}
    \includegraphics[width=0.49\linewidth]{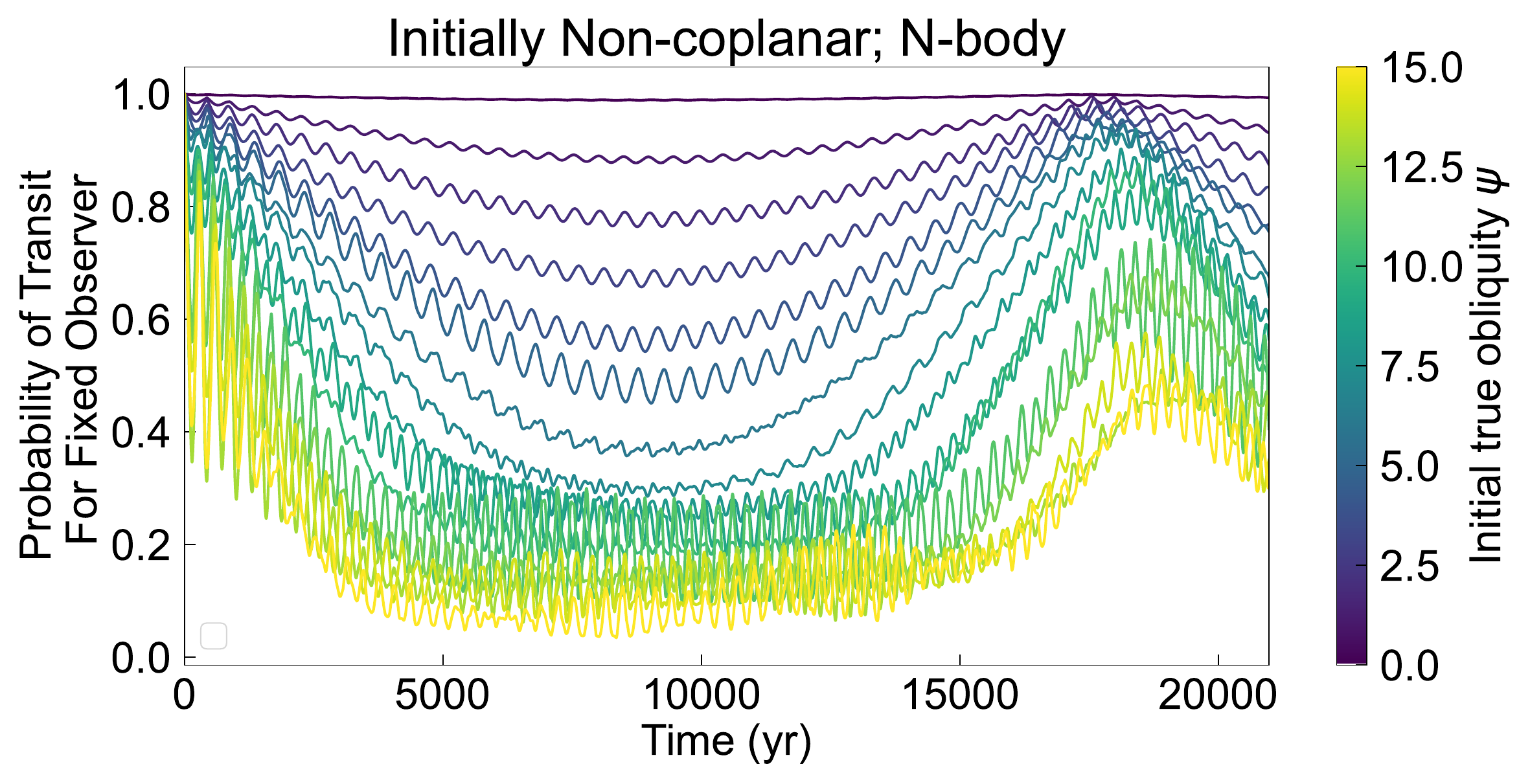}
    \caption{Same as Figure \ref{fig:mut_t_prob_fix}, the time-varying probability that an observer at a fixed position could detect both HD~93963~Ab and Ac transiting their host star, but obtained from N-body simulation as comparison.  The left and right panels are coplanar and non-coplanar initialization respectively. The color of the solid lines indicate the initial true obliquity.}
    \label{fig:mut_t_prob_fix_append}
\end{figure*}

Section \ref{sec:rm_res}: Figure \ref{fig:corner} illustrates the MCMC posterior distribution of RM+Transit+GP model. The results are summarized in Table \ref{tab:planet_para} in the main text. 

Section \ref{sec:dyn_ana_93963}: Figure \ref{fig:J2_variation_appendix} provides a more informative comparison of nodal precession varying by $J_2$ supplentary to Figure \ref{fig:J2_variation} in the main text. 

Section \ref{sec:obl-dyn}: Figure \ref{fig:mut_t_prob_fix_append} illustrates the time-varying probability that an observer at a fixed position could detect both HD~93963~Ab and Ac transiting their host star derived from N-body simulation as a comparison to the results derived from analytical solution (Figure \ref{fig:mut_t_prob_fix}, main text). 

\bibliography{sample631}{}
\bibliographystyle{aasjournal}

\end{document}